\documentclass[twocolumn]{aastex63}

\usepackage{fontawesome}
\usepackage{color}
\usepackage{graphicx}
\usepackage{amsmath, amssymb, bm}
\usepackage{CJK}
\usepackage[version=4]{mhchem} 
\usepackage{multirow}
\submitjournal{AAS Journals}

\shorttitle{The Hazy Atmosphere of GJ 1214 b}
\shortauthors{Gao et al.}

\begin{document}

\title{The Hazy and Metal-Rich Atmosphere of GJ 1214 b Constrained by Near and Mid-Infrared Transmission Spectroscopy}

\correspondingauthor{Peter Gao}
\email{pgao@carnegiescience.edu}

\author[0000-0002-8518-9601]{Peter Gao}
\affiliation{Earth \& Planets Laboratory, Carnegie Institution for Science, Washington, DC, USA}

\author[0000-0002-8518-9601]{Anjali A. A. Piette}
\affiliation{Earth \& Planets Laboratory, Carnegie Institution for Science, Washington, DC, USA}

\author[0000-0001-8342-1895]{Maria E. Steinrueck}
\affiliation{Max-Planck-Institut f\"ur Astronomie, Heidelberg, Germany}

\author[0000-0001-8236-5553]{Matthew C.\ Nixon}
\affiliation{Department of Astronomy, University of Maryland, College Park, MD, USA}

\author[0000-0002-0659-1783]{Michael Zhang}
\altaffiliation{51 Pegasi b fellow}
\affiliation{Department of Astronomy \& Astrophysics, University of Chicago, Chicago, IL, USA}

\author[0000-0002-1337-9051]{Eliza M.-R. Kempton}
\affiliation{Department of Astronomy, University of Maryland, College Park, MD, USA}

\author[0000-0003-4733-6532]{Jacob L.\ Bean}
\affiliation{Department of Astronomy \& Astrophysics, University of Chicago, Chicago, IL, USA}

\author[0000-0003-3963-9672]{Emily Rauscher}
\affiliation{Department of Astronomy, University of Michigan, Ann Arbor, MI, USA}

\author[0000-0001-9521-6258]{Vivien Parmentier}
\affiliation{Department of Physics, University of Oxford, Oxford, UK}
\affiliation{Universit\'e C\^ote d’Azur, Observatoire de la C\^ote d’Azur, CNRS, Laboratoire Lagrange, France}

\author[0000-0003-1240-6844]{Natasha E. Batalha}
\affiliation{NASA Ames Research Center, Moffett Field, CA, USA}

\author[0000-0002-2454-768X]{Arjun B.\ Savel}
\affiliation{Department of Astronomy, University of Maryland, College Park, MD, USA}
\affiliation{Center for Computational Astrophysics, Flatiron Institute, New York, NY, USA}

\author[0000-0002-3034-8505]{Kenneth E.\ Arnold}
\affiliation{Department of Astronomy, University of Maryland, College Park, MD, USA}

\author[0000-0001-8206-2165]{Michael T.\ Roman}
\affiliation{School of Physics and Astronomy, University of Leicester, Leicester, UK}

\author[0000-0003-0217-3880]{Isaac Malsky}
\affiliation{Department of Astronomy, University of Michigan, Ann Arbor, MI, USA}

\author[0000-0003-4844-9838]{Jake Taylor}
\affiliation{Department of Physics, University of Oxford, Oxford, UK}
\affiliation{Institut Trottier de Recherche sur les Exoplan\`etes and D\'epartement de Physique, Universit\'e de Montr\'eal, Montr\'eal, QC, Canada}

\begin{abstract}

The near-infrared transmission spectrum of the warm sub-Neptune exoplanet GJ 1214 b has been observed to be flat and featureless, implying a high metallicity atmosphere with abundant aerosols. Recent JWST MIRI LRS observations of a phase curve of GJ 1214 b showed that its transmission spectrum is flat out into the mid-infrared. In this paper, we use the combined near- and mid-infrared transmission spectrum of GJ 1214 b to constrain its atmospheric composition and aerosol properties. We generate a grid of photochemical haze models using an aerosol microphysics code for a number of background atmospheres spanning metallicities from 100 to 1000 $\times$ solar, as well as a steam atmosphere scenario. The flatness of the combined data set largely rules out atmospheric metallicities $\leq$300 $\times$ solar due to their large corresponding molecular feature amplitudes, preferring values $\geq$1000 $\times$ solar and column haze production rates $\geq$10$^{-10}$ g cm$^{-2}$ s$^{-1}$. The steam atmosphere scenario with similarly high haze production rates also exhibit sufficiently small molecular features to be consistent with the transmission spectrum. These compositions imply that atmospheric mean molecular weights $\geq$15 g mol$^{-1}$ are needed to fit the data. Our results suggest that haze production is highly efficient on GJ 1214 b and could involve non-hydrocarbon, non-nitrogen haze precursors. Further characterization of GJ 1214 b's atmosphere would likely require multiple transits and eclipses using JWST across the near and mid-infrared, potentially complemented by groundbased high resolution transmission spectroscopy.

\end{abstract}


\section{Introduction}\label{sec:intro}

The nature of the quintessential sub-Neptune GJ 1214 b has been a mystery since its discovery \citep{charbonneau2009}. Sitting on the larger-radii side of the radius gap \citep{vaneylen2018}, GJ 1214 b has been inferred to be either a gas dwarf or a water world \citep{millerricci2010,rogers2010gj1214b,nettelmann2011,valencia2013,luque2022,rogers2023arXiv}. In an effort to characterize its atmosphere and shed light on its interior, a large number of ground and space-based transmission spectroscopy programs have been conducted, focusing mostly on optical and near-infrared wavelengths due to instrumental limitations \cite[e.g.][]{bean2010,bean2011,crossfield2011,desert2011,berta2012,demooij2012,narita2013a,fraine2013,colon2013,teske2013,kreidberg2014,wilson2014}. However, to date all transmission spectra have been flat and featureless, or otherwise contaminated by stellar effects \citep{rackham2017}, suggesting the presence of high altitude aerosols and a high mean molecular weight atmosphere \citep{kreidberg2014}. 

Aerosol models that attempted to explain the flat transmission spectrum of GJ 1214 b can be roughly divided into two groups: those relying on clouds condensed out of trace gases in the atmosphere \citep{morley2013,morley2015,gao2018b,charnay2015b,ohno2018,ohno2020agg,christie2022} and those simulating hazes formed from photochemical reactions \citep{millerricci2012,morley2013,morley2015,adams2019,kawashima2019a,lavvas2019}. In general, the cloud models considered salt and sulfide clouds like \ce{KCl} and \ce{ZnS}, while the haze models used optical and physical properties of soots from combustion experiments and Titan tholins \citep{morley2015}. Both sets of models required high atmospheric metallicities, and high photochemical production rates in the case of hazes, to match the NIR data. These models also frequently predicted a decrease in aerosol opacity -- and thus the appearance of spectral features from atmospheric gases -- at longer wavelengths, owing to the small sizes needed for the aerosol particles to remain aloft at the high altitudes/low pressures suggested by the featureless NIR spectra. These models thus motivated observations at longer, mid-infrared wavelengths.




\citet{kempton2023} presented a 5-12 $\mu$m phase curve of GJ 1214 b measured by the JWST Mid-Infrared Instrument (MIRI) Low Resolution Spectrometer \citep[LRS;][]{kendrew2015}. The inferred day and nightside temperatures were 553 $\pm$ 9 K and 437 $\pm$ 19 K, respectively, which were significantly lower than GJ 1214 b's equilibrium temperature \citep[596 $\pm$ 19 K, assuming zero Bond albedo and full heat redistribution;][]{cloutier2021}, suggesting the presence of highly reflective aerosols. Free chemical retrievals of the day and nightside emission spectra showed tentative detections of water vapor. Comparison of the measured day and nightside emission spectra and the full phase curve to 3D general circulation models that included reflective hazes showed that a high mean molecular weight atmosphere, corresponding to metallicities of $\geq$100 $\times$ solar, are needed to explain the data.  

\begin{deluxetable*}{lcr}
\tablecolumns{3}
\tablecaption{Planetary and stellar parameters for GJ 1214 b used in this work. \label{table:parameters}}
\tablehead{
\colhead{Quantity} & \colhead{Value} & \colhead{Source}}
\startdata
Stellar effective temperature & 3250 K & \citet{kempton2023} \\
Stellar log(g) & 5.0 & \citet{kempton2023} \\
Stellar metallicity & +0.2 & \citet{kempton2023} \\
Stellar radius & 0.207 R$_{\sun}$ & See table footnote\tablenotemark{a} \\
Planet radius & 2.628 R$_{\earth}$\tablenotemark{b} & \citet{cloutier2021}; footnote\tablenotemark{a} \\
Planet mass & 8.17 M$_{\earth}$ & \citet{cloutier2021} \\
Planet intrinsic temperature & 30 K\tablenotemark{c} & \citet{lopez2014} \\
Planet semi-major axis & 0.01429 AU\tablenotemark{b} & \citet{cloutier2021}; footnote\tablenotemark{a}\\
\enddata
\tablenotetext{a}{We derive a new stellar radius for GJ 1214 by fitting the JWST MIRI LRS stellar spectrum with a PHOENIX model with the above stellar effective temperature, log(g), and [M/H]; the same model was used in the data reduction of \citet{kempton2023} and \citet{kreidberg2014}. }
\tablenotetext{b}{Derived using the planet-to-star radius ratio and semi-major axis-to-stellar radius ratio from \citet{cloutier2021} and the updated stellar radius above. We make the simplifying assumption that the planet-to-star radius ratio is the average of those measured in the 3.6 and 4.5 $\mu$m Spitzer channels \citep[see Table 4 of][]{cloutier2021}.}
\tablenotetext{c}{Estimated from Figure 5 of \citet{lopez2014}.}
\end{deluxetable*}

The transmission spectrum obtained from the phase curve measurements is flat and featureless, and at nearly the same band-integrated transit depth as the NIR data. A fit of the hazy GCM models to the combined NIR and MIRI transmission spectrum pointed to atmospheric metallicities $>$1000 $\times$ solar, but only limited haze properties were explored in those models \citep{kempton2023}. In this paper, we conduct a more in depth data-model comparison using a grid of haze microphysical models in order to ascertain the extent to which the NIR-to-MIR transmission spectrum of GJ 1214 b can constrain its atmospheric composition and haze properties.


In ${\S}$\ref{sec:model}, we describe how we generate our model grid. We compare our computed model transmission spectra to data in ${\S}$\ref{sec:results} and determine to what extent they constrain atmospheric composition and haze production rate. In ${\S}$\ref{sec:discussion} we discuss the implications of our results, what our models are missing, and what additional observational tests are needed to further improve our understanding of GJ 1214 b. We state our conclusions in ${\S}$\ref{sec:conclusions}. 

\section{Models}\label{sec:model}

\subsection{Modeling Strategy}\label{sec:strat}

In order to generate hazy model transmission spectra to compare to the observations, we make use of a series of 1-dimensional atmospheric models, with the outputs of one feeding into the next. To begin, we compute atmospheric temperature-pressure (TP) profiles using 
\texttt{GENESIS} \citep{gandhi2017,piette2020} and chemical abundance profiles using \texttt{FASTCHEM 2} \citep{stock2022}, assuming thermochemical equilibrium without rainout/condensation and a cloud/haze-free atmosphere, for a specific set of atmospheric compositions. The TP profiles are used to initialize the microphysical aerosol model \texttt{CARMA} \citep{turco1979,toon1988,jacobson1994,ackerman1995}, which simulates the particle size and vertical distribution of hazes. Assuming spherical particles, we then use a Mie code \citep[bhmie;][]{bohren2008} to generate the optical depth, single scattering albedo, and asymmetry parameter as a function of pressure level and wavelength from the particle distributions. Finally, the particle optical properties and TP and chemical abundance profiles are fed into \texttt{PICASO 3.0} \citep{mukherjee2023} to generate the transmission spectra. The planetary and stellar parameters we used in our modeling are given in Table \ref{table:parameters}.

\subsection{Background Atmosphere}\label{sec:thermal}

\begin{deluxetable}{lc}
\tablecolumns{2}
\tablecaption{Mean molecular weight (MMW) of model atmospheres at 1 bar.
 \label{table:mmw}}
\tablehead{
\colhead{Model Atmosphere} & \colhead{MMW (g mol$^{-1}$)}}
\startdata
100 $\times$ solar & 5.2  \\
300 $\times$ solar & 9.8 \\
500 $\times$ solar & 13.4 \\
1000 $\times$ solar & 19.5 \\
Steam & 18.0  \\
\enddata
\end{deluxetable}

\begin{deluxetable*}{lcccccc}
\tablecolumns{7}
\tablecaption{Parameters\tablenotemark{*} for defining atmospheric dynamic viscosity in units of Pa s using Equation \ref{eq:vis}.
 \label{table:vis}}
\tablehead{
\colhead{Species} & \colhead{C$_1$} & \colhead{C$_2$} & \colhead{C$_3$} & \colhead{C$_4$} & \colhead{T$_{min}$ (K)} & \colhead{T$_{max}$ (K)} }
\startdata
\ce{H2} & 1.7970 $\times$ 10$^{-7}$ & 0.685 & -0.59 & 140 & 13.95 & 3000 \\
\ce{He} & 3.2530 $\times$ 10$^{-7}$ & 0.7162 & -9.6 & 107 & 20 & 2000 \\
\ce{H2O} & 1.7096 $\times$ 10$^{-8}$ & 1.1146 & 0 & 0 & 273.16 & 1073.15 \\
\ce{CO} & 1.1127 $\times$ 10$^{-6}$ & 0.5338 & 94.7 & 0 & 68.15 &1250 \\
\ce{CO2} & 2.1480 $\times$ 10$^{-6}$ & 0.46 & 290 & 0 & 194.67 & 1500 \\
\ce{CH4} & 5.2546 $\times$ 10$^{-7}$ & 0.59006 & 105.67 & 0 & 90.69 & 1000 \\
\ce{N2} & 6.5592 $\times$ 10$^{-7}$ & 0.6081 & 54.714 & 0 & 63.15 & 1970 \\
\enddata
\tablenotetext{*}{All values from Table 2-138 of \citet{perry2018}.}
\end{deluxetable*}




\begin{figure*}[hbt!]
\centering
\includegraphics[width=0.8\textwidth]{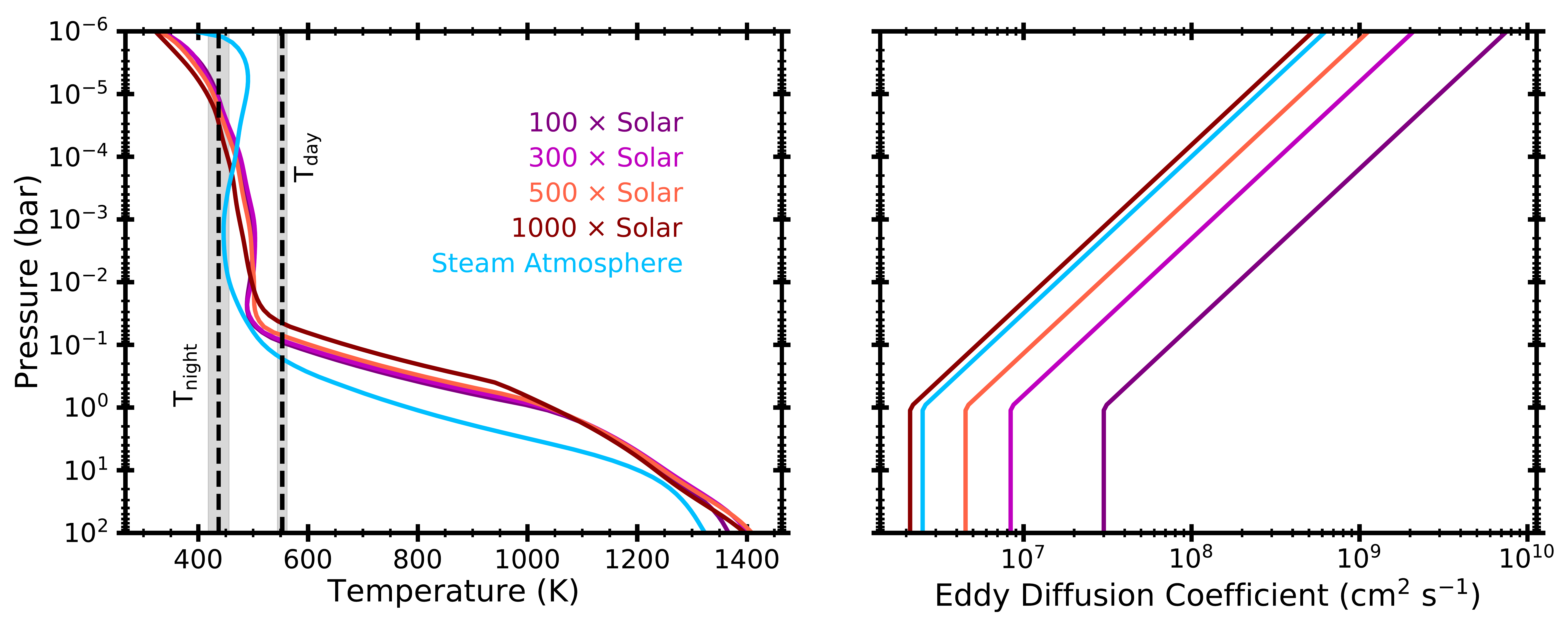}
\caption{Temperature-pressure (left) and eddy diffusion coefficient (right) profiles for the atmospheric compositions considered. The TP profiles were computed assuming radiative-convective equilibrium. The day and nightside temperatures of GJ 1214 b inferred by \citet{kempton2023} from the observed phase curve and their uncertainties are given in the dashed vertical lines and grey regions, respectively. }
\label{fig:tpkzz}
\end{figure*}

For the compositions of our background atmospheres, we consider N $\times$ solar metallicities, with N = 100, 300, 500, and 1000, in line with the findings of \citet{kempton2023}. We further examine a pure water vapor (steam) atmosphere motivated by the tentative detection of \ce{H2O} in the day and nightside emission spectra of GJ 1214 b \citep{kempton2023}. The atmospheric thermal structure for all of the considered background atmospheric cases, as well as the chemical abundance profiles of the N $\times$ solar metallicity cases are computed using the \texttt{GENESIS} and \texttt{FASTCHEM 2} codes. \texttt{GENESIS} is a self-consistent 1D atmospheric model that calculates the equilibrium TP profile and thermal emission spectrum under the assumptions of radiative-convective, hydrostatic and local thermodynamic equilibrium \citep{gandhi2017,piette2020,piette2020_tio}. \texttt{FASTCHEM 2} is a semi-analytical thermochemical equilibrium code \citep{stock2018,stock2022}. We couple \texttt{GENESIS} and \texttt{FASTCHEM 2} in order to iteratively solve for the equilibrium TP profile and chemical abundance profiles in parallel. Besides the TP profile, the inputs to \texttt{FASTCHEM 2} are the elemental abundances; we use the elemental abundances of \citet{asplund2009} with all elements but H and He enhanced by a factor of N for each N $\times$ solar metallicity case. \texttt{FASTCHEM 2} considers almost 500 chemical species in the calculation of thermochemical equilibrium. In the calculation of radiative transfer, however, we include only the primary opacity sources for the compositions explored: H$_2$O \citep{rothman2010}, CH$_4$ \citep{yurchenko2013,yurchenko2014}, C$_2$H$_2$ \citep{rothman2013,gordon2017}, CO$_2$ \citep{rothman2010}, CO \citep{rothman2010}, HCN \citep{harris2006}, NH$_3$ \citep{yurchenko2011}, N$_2$ \citep{barklem16,western18} and collision-induced absorption (CIA) due to H$_2$-H$_2$ and H$_2$-He \citep{richard2012}. The cross sections for these chemical species are calculated using the data cited above and the methods described in \citet{gandhi2017}. 

The energetic boundary conditions of the atmosphere are the incident stellar irradiation and the internal heat flux. For the stellar spectrum, we use a \texttt{PHOENIX} spectrum corresponding to $T_{\mathrm{eff}}$ = 3250\,K, log$(g) = 5.0$, and [M/H] = +0.2, as used in \citet{kreidberg2014} and \citet{kempton2023}. We use an internal heat flux equivalent to an internal temperature $T_{\rm int}$ = 30~K, which is estimated from Figure 5 of \citet{lopez2014} and is also representative of the values expected for GJ~1214~b for an age of 1--10~Gyr and a hydrogen-rich to water-rich atmosphere \citep{valencia2013}. We note, however, that the choice of $T_{\rm int}$ does not impact the temperature structure in the pressure range probed by the observations ($<$10 bars), as irradiation dominates over internal heat in this region. 

We further assume efficient heat redistribution from the day- to the nightside and a dayside Bond albedo of 0.6, similar to the albedo constraint of 0.51 $\pm$ 0.06 from \citet{kempton2023}. As discussed in the Methods section of \citet{kempton2023}, the derived Bond albedo of GJ 1214 b can actually vary between 0.39 and 0.61 when taking into account the full range of data reductions, and in particular the uncertainties in their measured nightside flux. Our choice of a Bond albedo on the high side of that range thus allows for a more conservative modeling strategy: the resulting cooler atmosphere leads to smaller molecular feature amplitudes and flatter transmission spectra for a given atmospheric composition, allowing for a wider range of compositions (i.e. lower atmospheric molecular weights) to be acceptable by the data constraints. We discuss the impact of temperature variations on our results in ${\S}$\ref{sec:atmsens}.

\begin{figure*}[hbt!]
\centering
\includegraphics[width=0.8\textwidth]{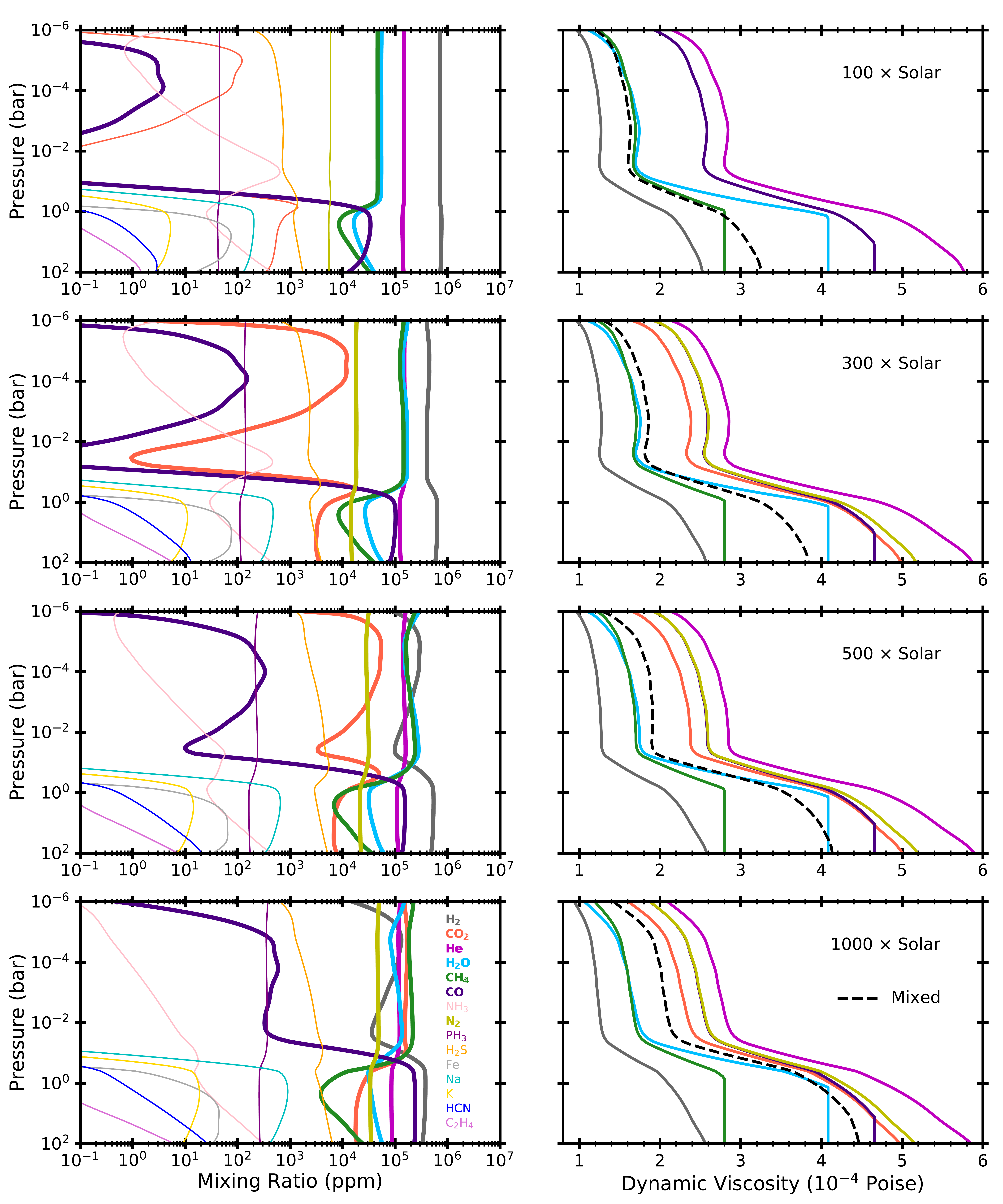}
\caption{Mixing ratio (left) and atmospheric dynamic viscosity (right) profiles for the N $\times$ solar metallicity cases. The mixing ratio profiles of the species considered in the viscosity mixing calculation are shown as thicker curves and their labels are bolded. The mixed dynamic viscosity profiles are shown in the dashed curves. The constant-with-pressure viscosities for several species at higher pressures are due to the high temperature limits for the viscosity expressions of those species (Table \ref{table:vis}). }
\label{fig:mrvis}
\end{figure*}

Figure \ref{fig:tpkzz} shows the computed TP profiles and Figure \ref{fig:mrvis} shows the mixing ratio profiles of chemical species with mixing ratios of at least 1 ppm at some point in the atmosphere for the N $\times$ solar cases. The temperatures near the photospheres of all cases lie between the inferred day and nightside temperatures of GJ 1214 b from \citet{kempton2023}, appropriate for our modeling of the transmission spectrum at the limb. \ce{H2}, \ce{He}, \ce{H2O}, \ce{CH4}, and \ce{N2} are consistently the most abundant species across all of these cases at pressures $<$1 bar, with \ce{CO} and \ce{CO2} becoming comparatively abundant at higher ($\geq$500 $\times$ solar) metallicities, consistent with previous studies \citep{moses2013c,hu2014,morley2015,kawashima2019b,lavvas2019}.

\begin{figure}[hbt!]
\centering
\includegraphics[width=0.45\textwidth]{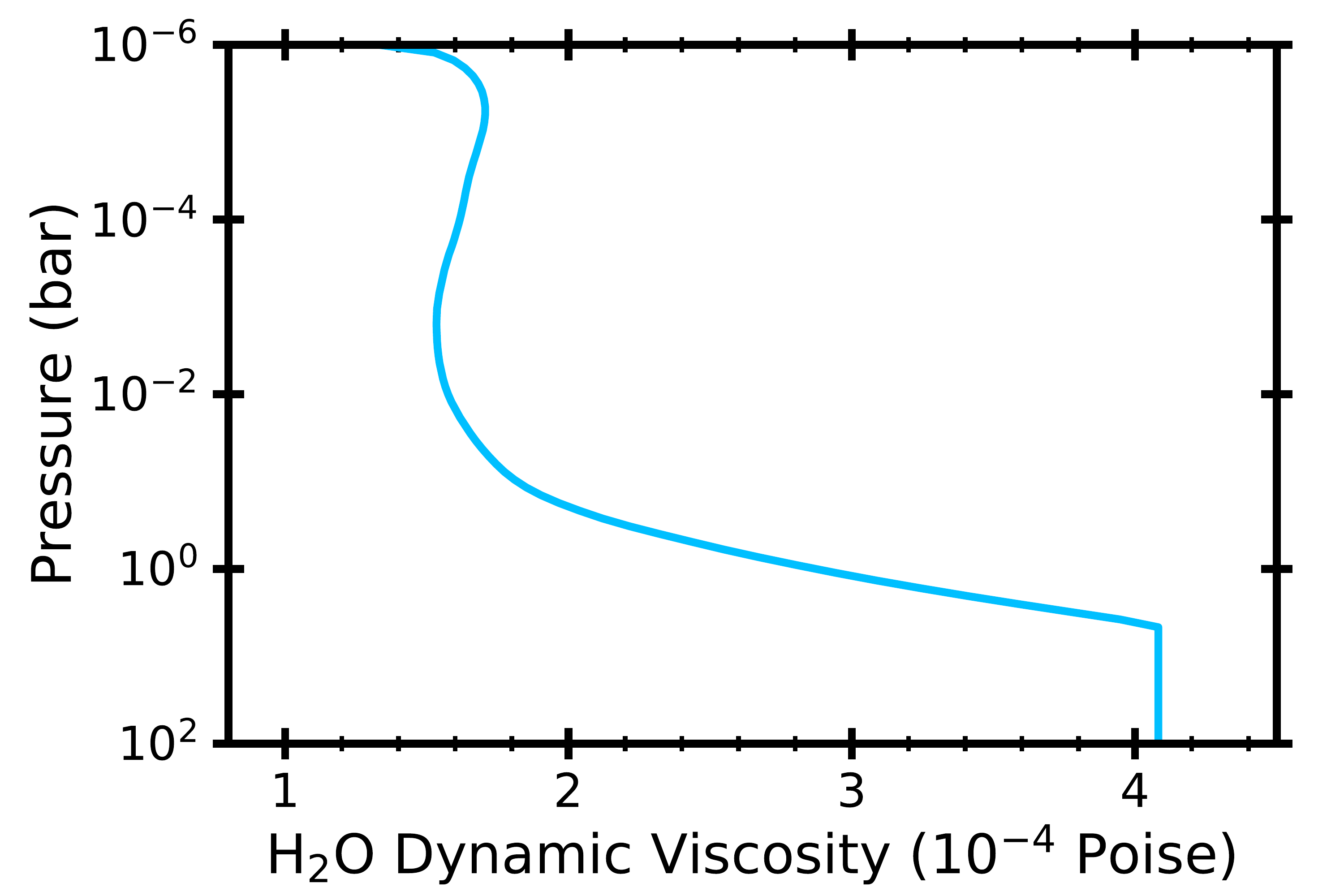}
\caption{Atmospheric dynamic viscosity for water vapor/the steam atmosphere case. The constant-with-pressure viscosity at pressures greater than a few bars is due to the high temperature limit for water vapor's viscosity expression (Table \ref{table:vis}). }
\label{fig:h2otitanvis}
\end{figure}

We use the eddy diffusion approximation to simulate vertical mixing of haze particles. We consider the eddy diffusion coefficient ($K_{zz}$) profiles derived by \citet{charnay2015a} from GCM simulations with tracers:

\begin{equation}
\label{eq:contangle}
K_{zz} = K_{zz0}P^{-0.4}
\end{equation}

\noindent where $P$ is the pressure in bars and $K_{zz0}$ is a constant. We assume that the $K_{zz}$ profile is constant at pressures $>$1 bar with a value of $K_{zz0}$ to avoid numerical issues; this does not impact our results since transmission spectroscopy probes much lower pressures, especially for hazy atmospheres \citep{fortney2003,morley2015,gao2020sp}. Figure \ref{fig:tpkzz} shows the $K_{zz}$ profiles we used for our different model atmospheres. Here we make the simplifying assumption that, for all of our models, $K_{zz0}$ $\propto$ $H^2$, with $H$ being the pressure scale height, such that the eddy mixing timescale $H^2/K_{zz0}$ is constant with changes in $H$, which is mostly due to variations in the atmospheric mean molecular weight for our different atmospheric composition cases (Table \ref{table:mmw}). We scale using $H$ from the 100 $\times$ solar model at 1 bar, for which we set $K_{zz0}$ to 3 $\times$ 10$^7$ following \citet{charnay2015a}. Our scaling allows for similar magnitudes of mixing for different model atmospheres, such that we can ignore variations thereof as a contributing factor to differences in the haze distribution across our models. Interestingly, in \citet{charnay2015a}, $K_{zz0}$ is 3 $\times$ 10$^6$ cm$^2$ s$^{-1}$ for the pure steam atmosphere, which has a mean molecular weight $\sim$3 times that of the 100 $\times$ solar atmosphere, following our scaling closely. However, at lower metallicities the scaling breaks down, i.e. $K_{zz0}$ = 7 $\times$ 10$^6$ cm$^2$ s$^{-1}$ for the solar metallicity case of \citet{charnay2015a}. Indeed, the actual variation in atmospheric mixing with changing atmospheric composition is doubtlessly more complex \citep{kataria2014,zhang2017comp,drummond2018met}, and future studies are needed to quantify these dependencies at very high metallicities.






Sedimentation of haze particles is a major control of the haze distribution, particularly at lower pressures \citep{parmentier2013,steinrueck2021}. The sedimentation velocity is dependent upon the atmospheric dynamic viscosity, which is a function of the atmospheric composition. As our model atmospheres are dominated by multiple species, we must compute a ``mixed'' viscosity profile from those of the individual species. Here we use the method of \citet{davidson1993} that accounts for the efficiency of momentum transfer between gas molecules, where the mixed viscosity $\eta_{mix}$ is given by

\begin{equation}
\label{eq:mixvis}
\eta_{mix} = \left ( \sum_{i,j} \frac{y_i y_j}{\sqrt{\eta_i \eta_j}} E_{i,j}^{3/8} \right )^{-1}
\end{equation}

\noindent where $\eta_i$ is the viscosity of gas $i$, $y_i$ is the momentum fraction of gas $i$ defined by 

\begin{equation}
\label{eq:mfrac}
y_i =  \frac{x_i \sqrt{M_i}}{\sum_i x_i \sqrt{M_i}} 
\end{equation}

\noindent with $x_i$ and $M_i$ as the mixing ratio and molecular weight of gas $i$, respectively. Finally, $E_{i,j}$ is the efficiency of momentum transfer between gases $i$ and $j$ given by

\begin{equation}
\label{eq:eff}
E_{i,j} =  \frac{2\sqrt{M_i M_j}}{M_i + M_j} 
\end{equation}

\noindent The summation in Equation \ref{eq:mixvis} is taken over each pair of gases in the atmosphere. For simplicity, we only consider gases with mixing ratios $>$1\% in our models for the viscosity mixing calculation, which includes \ce{H2}, \ce{He}, \ce{CO}, \ce{H2O}, and \ce{CH4} for the 100 $\times$ solar case, and the addition of \ce{CO2} and \ce{N2} for the higher metallicity cases. The steam atmosphere case consists of just the \ce{H2O} viscosity. The dynamic viscosity of each individual gas $i$ is parameterized using 

\begin{equation}
\label{eq:vis}
\eta_i = \frac{C_1 T^{C_2}}{1+C_3/T+C_4/T^2}
\end{equation}

\noindent where $T$ is the temperature in units of K and the constants $C_1$, $C_2$, $C_3$, and $C_4$ are given in Table \ref{table:vis}. Figure \ref{fig:mrvis} shows the viscosity profiles of the individual gases considered and the mixed viscosity profile for the N $\times$ solar cases, while Figure \ref{fig:h2otitanvis} shows the viscosity profile for the steam atmosphere. Note that several of the viscosity parameterizations possess maximum valid temperatures lower than the deep temperature of our atmospheric models; in such cases, we set the viscosity at higher temperatures to those at the maximum valid temperature. This assumption should not affect our results, as at such high pressures atmospheric mixing dominates over sedimentation in terms of particle transport processes. 

\subsection{Haze Microphysics}\label{sec:micro}

The Community Aerosol and Radiation Model for Atmospheres (\texttt{CARMA}) is an aerosol microphysics code that solves the aerosol continuity equation taking into account particle nucleation, condensation, evaporation, coagulation, sedimentation, diffusion, and advection \citep[see][for a complete description of the model]{gao2018a}. It relies on a bin scheme such that the particle size distribution is fully resolved instead of assuming a functional form. For hazes, we only consider coagulation and transport, as the initial formation process of hazes through chemical reactions and, potentially, nucleation/condensation, is highly uncertain \citep{horst2018b,gao2021rev}.


For each model atmosphere, we simulate haze production as a downward flux of spherical particles from the top of the atmosphere at 1 $\mu$bar. We consider column haze production rates of 10$^{-14}$, 10$^{-13}$, 10$^{-12}$, 10$^{-11}$, 10$^{-10}$, and 10$^{-9}$ g cm$^{-2}$ s$^{-1}$. The lower end of our range is motivated by the haze production rate at Titan \citep{checlair2016}, while the higher end of the range stems from estimates from GJ 1214 b photochemical models \citep{kawashima2019b,lavvas2019}. We assume an initial particle size of 10 nm and a mass density of 1 g cm$^{-3}$, similar to those in the upper atmosphere of Titan \citep{lavvas2010}. The particles are allowed to coagulate with each other as they sediment and mix into the lower atmosphere, generating larger spherical particles with a sticking efficiency of 1. We do not simulate porous fractal aggregates in this work for simplicity, though we discuss the implications of such particles in ${\S}$\ref{sec:hazesens}. No destruction process is modeled in our work, as the thermal decomposition pressure/temperature of exoplanet hazes is uncertain \citep{lavvas2017}. This does not affect our results unless the hazes are destroyed at sufficiency low pressures/temperatures such that they are optically thin. 

The single scattering albedo of the aerosols needed to reproduce the day and nightside emission spectrum of GJ 1214 b lies between $\sim$1, i.e.\ purely scattering, and those of Titan tholins \citep{kempton2023}. As such, we consider both compositions in this work as endmembers. The complex refractive indices for Titan tholins were sourced from \citet{khare1984}. For the purely scattering haze particles, we assume a simplified, wavelength-independent complex refractive index of $m$ = 1.8 + 10$^{-9}i$, where 1.8 is the value of the real refractive index for soots at 1.4 $\mu$m, motivated by the use of the soot real refractive indices for the purely scattering haze in the general circulation models of \citet{kempton2023}. The imaginary refractive index is chosen to be sufficiently low as to allow for maximum scattering without incurring numerical issues with the Mie calculations. A purely scattering haze would exhibit a Rayleigh-scattering slope (i.e. $\lambda^{-4}$) at wavelengths much greater than the particle size, while a more absorptive haze would result in a shallower slope and more absorption at longer wavelengths. Therefore, a purely scattering haze would be the most difficult type of haze with which to generate a flat transmission spectrum out to mid-IR wavelengths. While recent laboratory experiments have generated haze refractive indices more appropriate to GJ 1214 b's atmospheric conditions \citep{corrales2023,he2023arXiv}, we will restrict our study to the above two cases following \citet{kempton2023}, and look to future studies to evaluate the impact of the new refractive indices on GJ 1214 b model transmission spectra. 

\begin{figure*}[hbt!]
\centering
\includegraphics[width=0.8\textwidth]{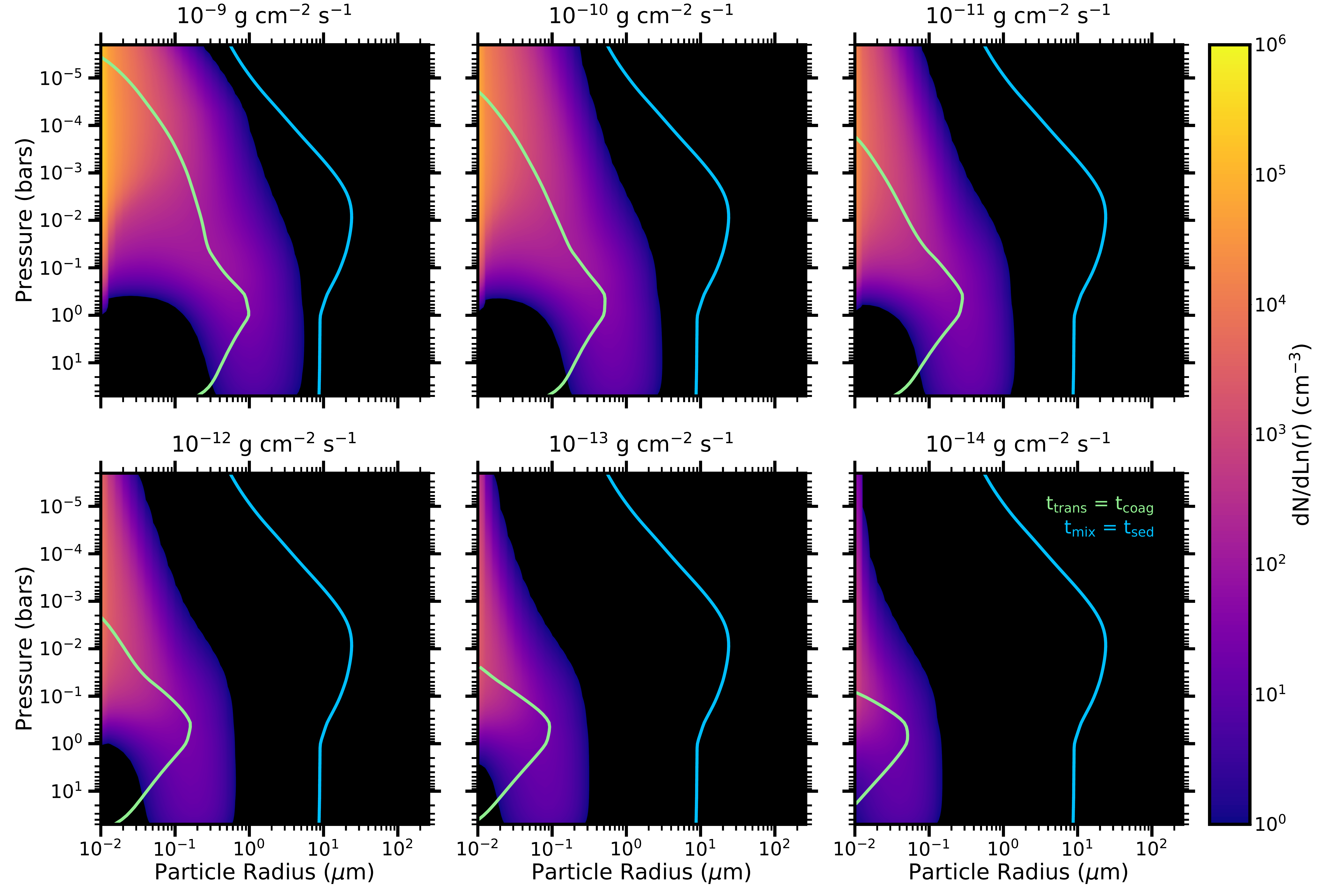}
\caption{Number density of haze particles as a function of particle radius and pressure level in the atmosphere for the 1000 $\times$ solar metallicity case and various haze production rates. The blue and green curves indicate particle radii where the eddy mixing timescale is equal to the sedimentation timescale and where the transport timescale (set by the smaller of the eddy mixing and sedimentation timescales) equal the coagulation timescale, respectively. See \citet{gao2020sp} for our definition of these timescales. Particles smaller than the green curve coagulate faster than can be transported, and vice versa. Particle transport is dominated by mixing for particles smaller than the blue curve, and sedimentation for particles that are larger. }
\label{fig:hazedist}
\end{figure*}

\begin{figure*}[hbt!]
\centering
\includegraphics[width=0.8\textwidth]{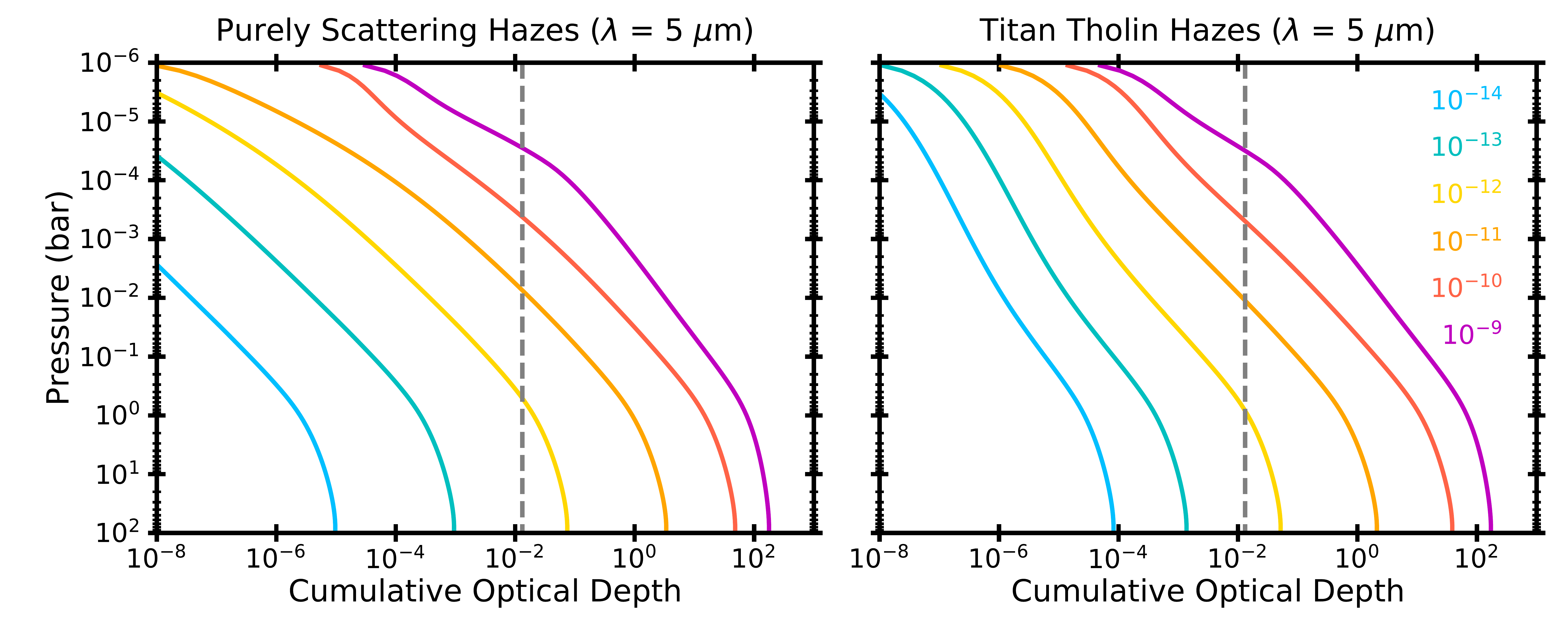}
\caption{Nadir column optical depth at a wavelength of 5 $\mu$m for purely scattering (left) and Titan tholin (right) hazes for the 1000 $\times$ solar metallicity case and various haze production rates. The vertical gray dashed line indicates the nadir column optical depth where the slant optical depth equals 1 for the geometry of GJ 1214 b's atmosphere, given the updated planetary parameters (Table \ref{table:parameters}) and mean molecular weight of a 1000 $\times$ solar metallicity atmosphere (Table \ref{table:mmw}). }
\label{fig:opdcuml}
\end{figure*}

\begin{figure*}[hbt!]
\centering
\includegraphics[width=0.8\textwidth]{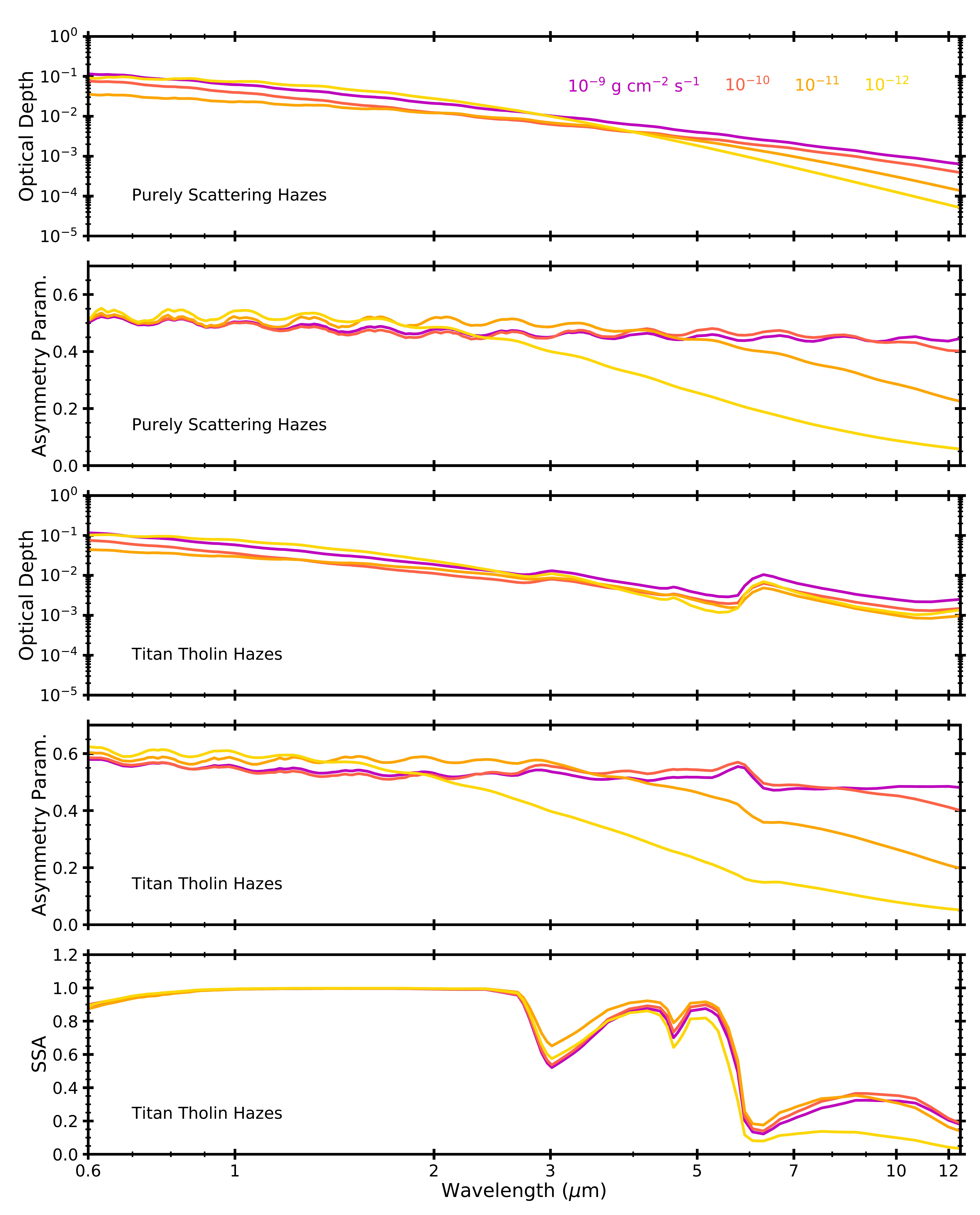}
\caption{The optical properties of the haze models shown in Figure \ref{fig:opdcuml} at a slant optical depth at 5 $\mu$m of 1 (i.e. where the colored curves intersect the gray dashed line), as a function of wavelength. Models with haze production rates $<$10$^{-12}$ g cm$^{-2}$ s$^{-1}$ are not shown as they never achieve a slant optical depth at 5 $\mu$m of 1 in the atmosphere. The top two plots show the optical depth and asymmetry parameter (``Asymmetry Param.'') of the purely scattering hazes while the bottom three plots show the optical depth, asymmetry parameter, and single scattering albedo (``SSA'') of the Titan tholin hazes. We do not show the single scattering albedo of the purely scattering hazes because it is essentially 1 for the wavelength range of interest. Minor wave-like features in the asymmetry parameter and optical depth curves are due to numerical errors in the Mie calculation. }
\label{fig:opticalprops}
\end{figure*}

\subsection{Transmission Spectra}\label{sec:spec}

\texttt{PICASO 3.0} is the newest version of the \texttt{PICASO} code \citep{batalha2017,mukherjee2023} that now includes the capability of computing the planetary transmission spectrum. The methodology of the transit code relies on computing the slant optical depth and integrating the slant transmittance following Eq. 11 of \citet{Brown2001} \footnote{\href{https://github.com/natashabatalha/picaso/blob/9383dc0f4a97e8fafe0834fec6f4e1f31ce5b6a0/picaso/fluxes.py}{\faCode picaso.fluxes.get\_transit\_1d()}}.
 The optical depths are computed from the cross section methodology outlined in \citet{freedman2014} and \citet{GharibNezhad2021}. Specific to this analysis, we use the following line list databases in our cross section calculations: \ce{H2O} \citep{Polyansky2018H2O}, CO \citep{HITEMP2010, HITRAN2016,li15rovibrational}, \ce{CO2} \citep{HUANG2014reliable}, and \ce{CH4} \citep{yurchenko13vibrational, yurchenko_2014}. We include collision-induced-absorption from \ce{H2-H2}, \ce{H2-He}, \ce{H2-CH4}, and \ce{H2-N2} from \citep{saumon2012}. Lastly, \texttt{PICASO} self-consistently computes the Rayleigh cross-section of each scattering species using the King correction factor and polarisability of each molecule\footnote{\href{https://github.com/natashabatalha/picaso/blob/master/picaso/rayleigh.py}{\faCode 
 picaso.rayleigh.Rayleigh()}}, following the methodology described in Section 3.2.2 of \citet{MacDonald2022ApJ}. 
The transit functionality of \texttt{PICASO} was benchmarked with the comparisons to three independent codes (\texttt{ATMOS}, \texttt{PHOENIX}, and \texttt{CHIMERA}) in \citet{2022arXiv220811692T}.

\section{Results}\label{sec:results}

\subsection{Haze Distributions and Optical Properties}\label{sec:hazeprop}

The vertical and size distributions of haze particles vary substantially across the parameter space we explore here. These variations are caused by the order of magnitude changes in the haze production rate, as shown in Figure \ref{fig:hazedist} for the 1000 $\times$ solar metallicity case. In general, the particle size distribution evolves from a ``shoulder'' of the 10 nm seed particles in the upper atmosphere to a larger ``coagulation mode'' at depth. Higher haze production rates generate higher number densities and larger particles ($\sim$0.1-1 $\mu$m in the upper atmosphere and at depth) due to rapid growth by coagulation, while lower haze production rates result in little growth beyond the 10 nm seeds. The results for the other background atmosphere cases are similar. 

The peak of the haze size distribution at a given pressure level is set by where the coagulation timescale equals the vertical transport timescale, as represented by the green curve in Figure \ref{fig:hazedist}: when particles are small and numerous (i.e. when they are on the left side of the green curve), coagulation occurs fast enough to overwhelm any particle transport due to the coagulation rate being proportional to particle number density, leading to rapid localized particle growth; however, as coagulation is mass-conserving, it causes the number density of particles to decrease while increasing the particle sizes, leading to a reduction in the coagulation rate until transport becomes dominant, removing the particles from the local gas parcel and quenching further growth. The primary mode of particle transport for nearly all particles is eddy mixing, as most of the size distributions lie to the left of the blue curve in Figure \ref{fig:hazedist} where mixing dominates over sedimentation. At depth, coagulation slows due to the loss of small particles to coagulation, leading to the peak in the green curve at $\sim$0.1-1 bar. 

Higher haze production rates lead to larger haze optical depths due to the increased particle sizes and number densities (Figure \ref{fig:opdcuml}). At the highest haze production rates, the haze becomes optically thick in the transit (slant) geometry at pressures as low as 10 $\mu$bar, for both purely scattering and Titan tholin hazes. In contrast, at low haze production rates ($<$10$^{-12}$ g cm$^{-2}$ s$^{-1}$), the haze may not reach slant optical depth of 1 at all (at a wavelength of 5 $\mu$m). Here we define the ratio of slant to nadir optical depth $\tau_s/\tau_n$ as \citep{fortney2005},

\begin{equation}
\label{eq:slant}
\frac{\tau_s}{\tau_n} = \sqrt{\frac{2\pi R_p}{H}}
\end{equation}

\noindent where $R_p$ is the planet radius, which we take to be the value in Table \ref{table:parameters}. Defining the scale height $H$ using a temperature of 495 K \citep[i.e. midway between the measured day and nightside temperatures;][]{kempton2023}, gravity computed from the planetary parameters in Table \ref{table:parameters}, and an atmospheric mean molecular weight of 19.5 g mol$^{-1}$ for the 1000 $\times$ solar metallicity case (Table \ref{table:mmw}), we find $H$ $\sim$ 18 km and $\tau_s/\tau_n$ $\sim$ 76. In other words, the slant optical depth reaches 1 when the nadir optical depth reaches 1/76 $\sim$ 0.013. In general, we see column optical depth decreasing faster with decreasing haze production rate for the purely scattering haze compared to the Titan tholin haze due to the increased absorption of the latter cases. 

At the pressure levels where the haze becomes optically thick in the slant direction, we find varying behaviors in the haze optical properties as a function of haze production rate due to differences in the mean particle sizes and haze compositions (Figure \ref{fig:opticalprops}). By design, the purely scattering hazes possess single scattering albedo of nearly 1 with no spectral features at any wavelengths. However, the spectral slope and scattering properties can still be variable due to particle size differences. For example, the asymmetry parameter tends to stay fairly constant at shorter wavelengths but reduces to near zero at longer wavelengths once the ratio of particle size to wavelength is sufficiently small such that the particle becomes a Rayleigh scatterer. Meanwhile, Titan tholin hazes possess spectral absorption features at 3, 4.5, and 6.5 $\mu$m owing to vibrational modes of organic bonds \citep{wakeford2015}, which are visible in the optical depth curves and show up as low values in the single scattering albedo. Minor wave-like features in the asymmetry parameter and optical depth curves are due to numerical errors in the Mie calculation.

\subsection{Comparisons to Data}

We compare our model transmission spectra to three data sets spanning the near and mid-infrared: the HST WFC3 transmission spectrum from \citet{kreidberg2014}, the JWST MIRI LRS transmission spectrum from \citet{kempton2023}, and the Spitzer 3.6 and 4.5 $\mu$m photometric points from \citet{fraine2013}. For the latter, we specifically compare to the transit depths derived using orbital parameters from \citet{berta2012} in order to maximize consistency in orbital parameters between the three data sets. We fit the model spectra to the data using \texttt{scipy.optimize.minimize} to shift the models up and down until the reduced chi squared ($\chi_r^2$) is minimized. As the radius of the planet is known, shifting the model transit depths implies changing the reference pressure of the planet radius. However, changing the reference pressure has subtle effects, such as on the profile of gravity in the atmosphere, that is not captured by simply shifting the model spectra up and down. As such, we pick the reference pressure for each spectrum such that the shifts are never more than 30 ppm in either direction, in keeping with the uncertainties of the most precise data points \citep[i.e. the HST data from][]{kreidberg2014} Here we consider four degrees of freedom for calculating $\chi_r^2$: the shifts, the atmospheric composition, the haze production rate, and the haze composition (purely scattering haze vs. Titan tholin haze). We do not consider the MIRI LRS points $>$10 $\mu$m in the fitting due to uncertainties in the background subtraction at these wavelengths \citep[see][for details]{kempton2023}.

\begin{figure*}[hbt!]
\centering
\includegraphics[width=0.8 \textwidth]{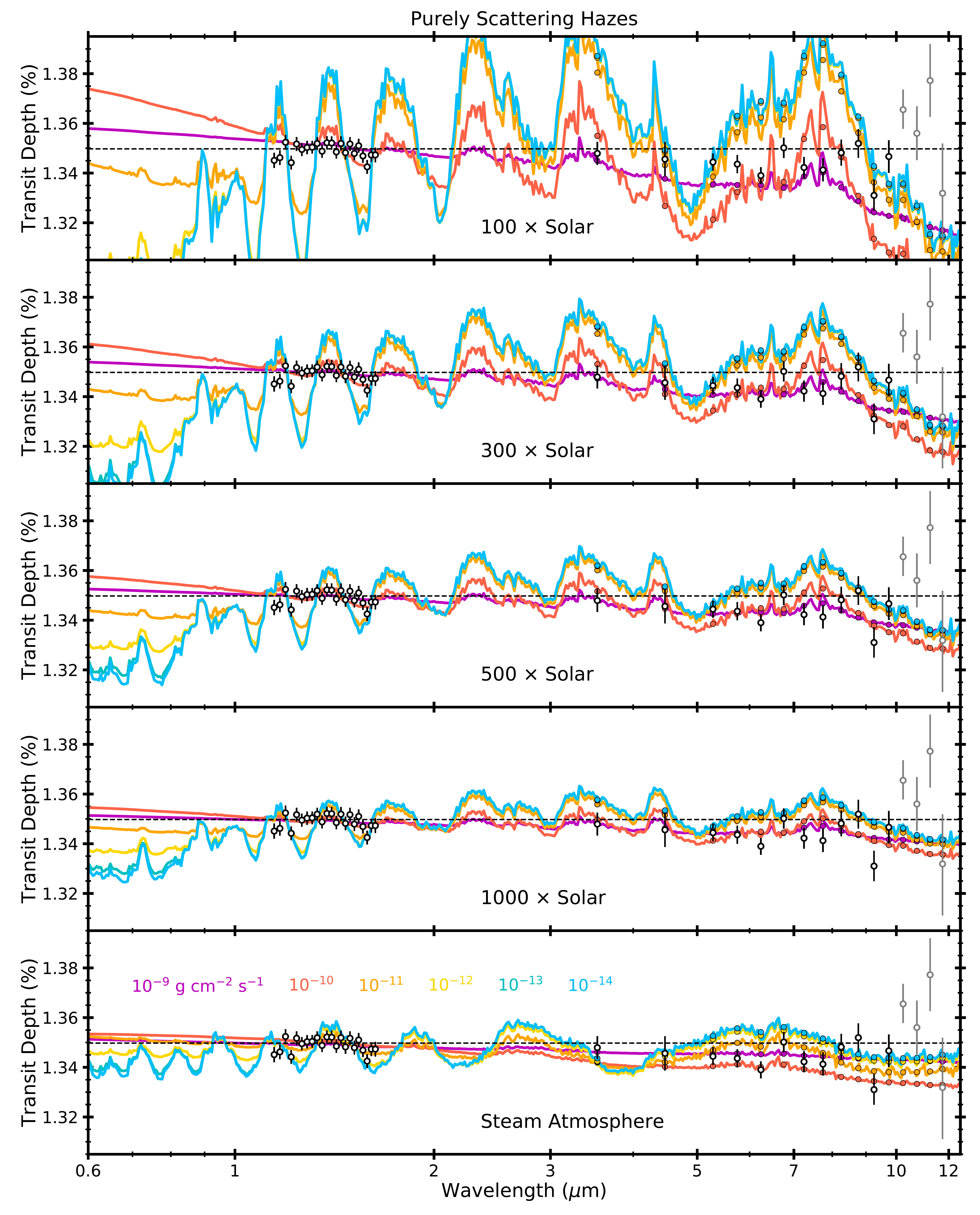}
\caption{Comparison of model transmission spectra for the purely scattering haze cases and various atmospheric compositions and haze production rates with the HST WFC3 \citep{kreidberg2014}, Spitzer \citep{fraine2013}, and JWST MIRI LRS \citep{kempton2023} observations (white points with black edges and errorbars; those with gray edges and errorbars were not considered in the fitting; see text). The band-integrated model transmission spectrum for the corresponding data points are shown in the colored points. The best fit flat line is shown in the black dashed line.}
\label{fig:spec}
\end{figure*}

\begin{figure*}[hbt!]
\centering
\includegraphics[width=0.8 \textwidth]{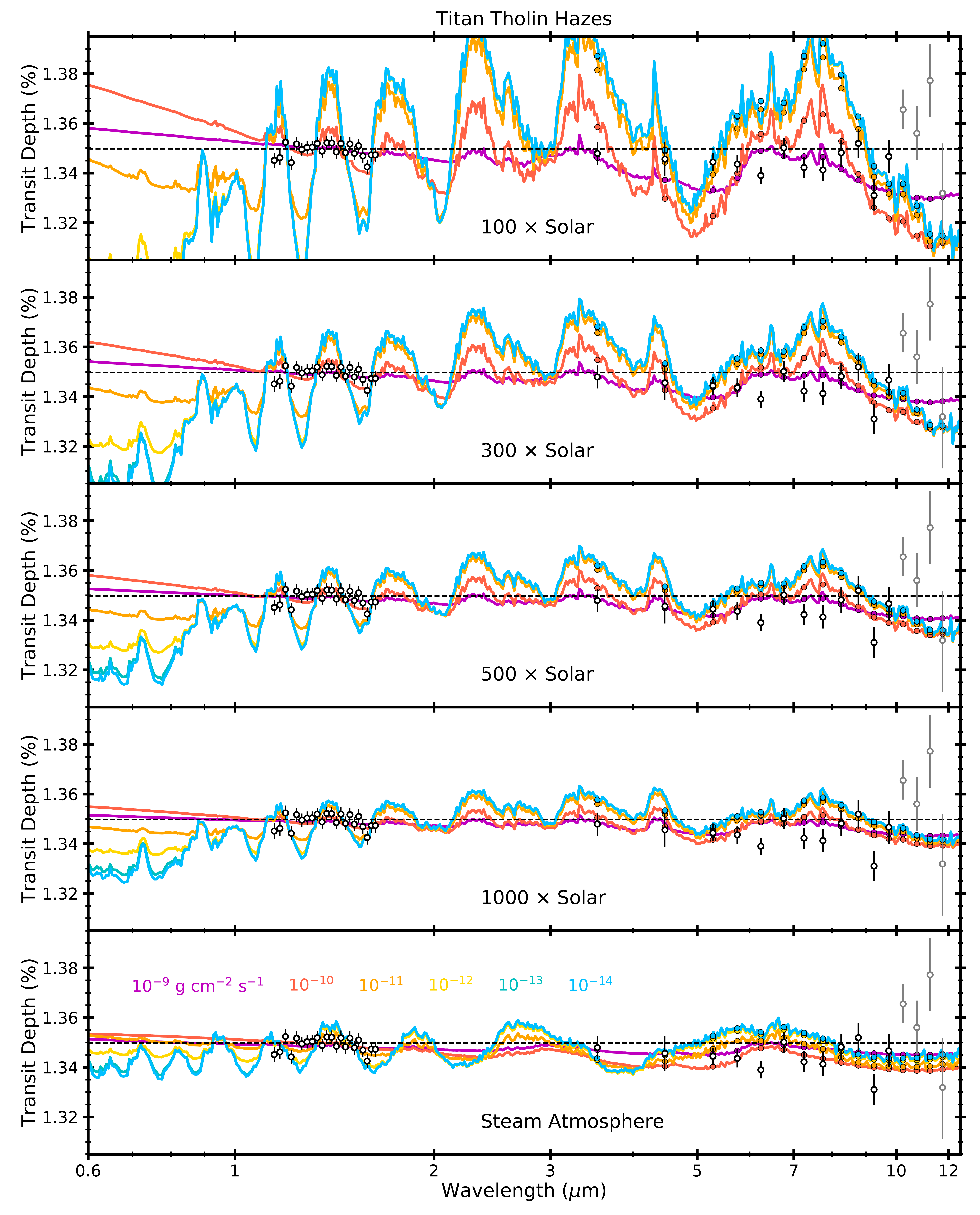}
\caption{Same as Figure \ref{fig:spec}, but for the Titan tholin haze cases.  }
\label{fig:specthol}
\end{figure*}

\begin{figure*}[hbt!]
\centering
\includegraphics[width=0.8 \textwidth]{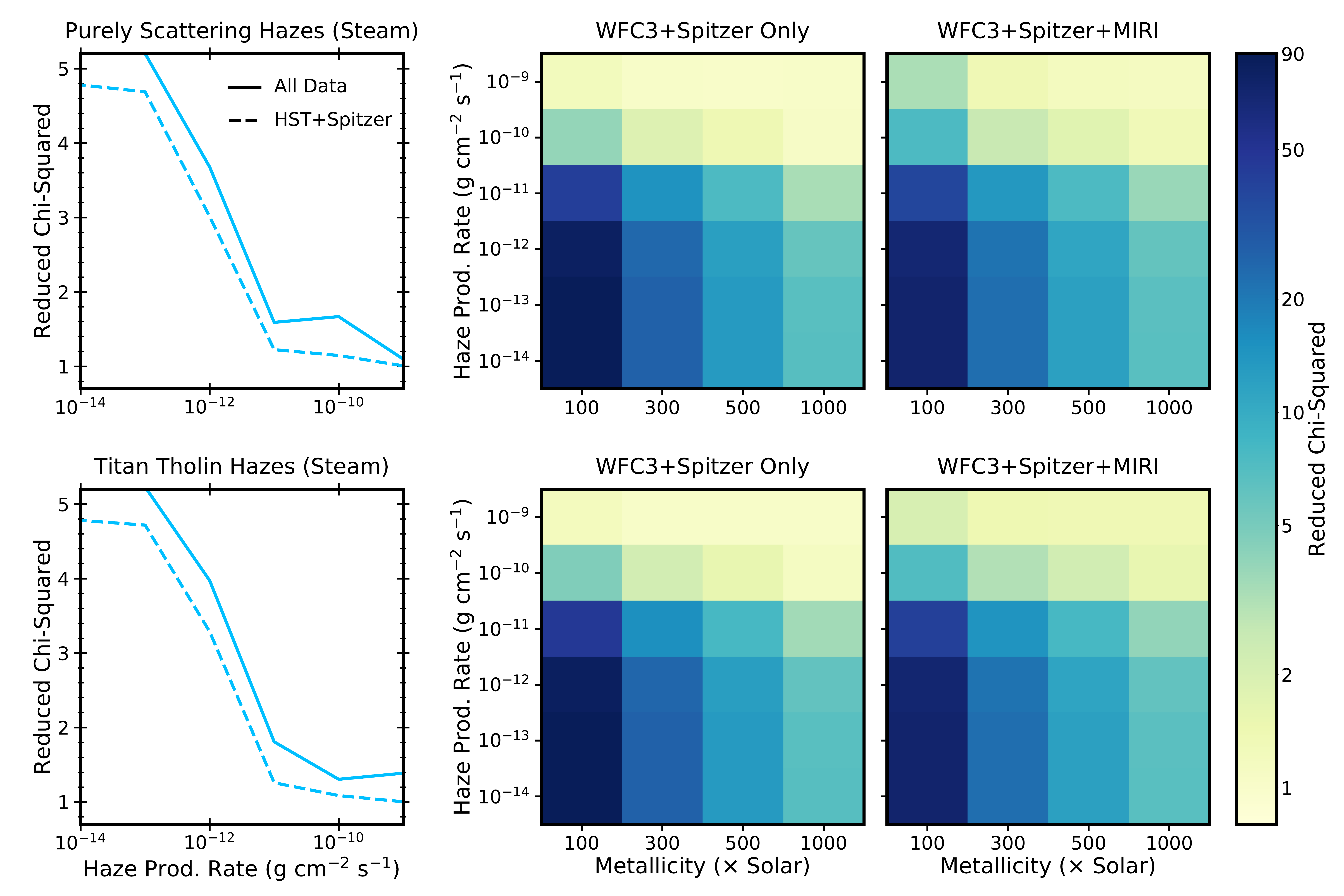}
\caption{Reduced chi-squared ($\chi_r^2$) values for the data-model comparisons for the steam atmosphere cases with (solid) and without (dashed) JWST MIRI LRS data (left) and the N $\times$ solar cases without (middle) and with (right) JWST MIRI LRS data. The top row shows results for the purely scattering haze cases, while the bottom row shows the same for the Titan tholin haze cases.}
\label{fig:rcs}
\end{figure*}

\begin{deluxetable*}{lccccc}
\tablecolumns{6}
\tablecaption{$\chi_r^2$ values for our data-model comparisons\tablenotemark{*}. 
 \label{table:rcs}}
\tablehead{
 & & \colhead{Pure Scattering} \vspace{-0.37cm} & \colhead{Pure Scattering} & \colhead{Titan Tholin} & \colhead{Titan Tholin} \\   \colhead{Atmo. Composition} \vspace{-0.37cm} & \colhead{Haze Prod. Rate (g cm$^{-2}$ s$^{-1}$)} & & & & \\ &  & \colhead{$\chi_r^2$ (No MIRI)} & \colhead{$\chi_r^2$ (All Data)} & \colhead{$\chi_r^2$ (No MIRI) } & \colhead{$\chi_r^2$ (All Data)}  }
\startdata
\multirow{ 6}{*}{100 $\times$ Solar} & 10$^{-9}$ & 1.22 & 3.31 & 1.2 & 2.03 \\
 & 10$^{-10}$ & 3.96 & 7.55 & 4.66 & 7.31 \\
 & 10$^{-11}$ & 42.49 & 38.01 & 46.35 & 41.82 \\
 & 10$^{-12}$ & 81.62 & 68.62 & 83.18 & 69.9 \\
 & 10$^{-13}$ & 87.83 & 73.31 & 87.89 & 73.36 \\
 & 10$^{-14}$ & 88.01 & 73.44 & 88.01 & 73.44 \\
\hline
\multirow{ 6}{*}{300 $\times$ Solar} & 10$^{-9}$ & 1.03 & 1.35 & 1.03 & 1.4 \\
 & 10$^{-10}$ & 1.84 & 2.53 & 2.2 & 3.1 \\
 & 10$^{-11}$ & 14.59 & 13.49 & 15.56 & 14.46 \\
 & 10$^{-12}$ & 24.64 & 21.39 & 25.02 & 21.7 \\
 & 10$^{-13}$ & 26.34 & 22.68 & 26.37 & 22.7 \\
 & 10$^{-14}$ & 26.41 & 22.73 & 26.41 & 22.73 \\
\hline
\multirow{ 6}{*}{500 $\times$ Solar} & 10$^{-9}$ & 1.01 & 1.16 & 1.02 & 1.34 \\
 & 10$^{-10}$ & 1.37 & 1.74 & 1.57 & 2.21 \\
 & 10$^{-11}$ & 7.59 & 7.48 & 7.99 & 7.91 \\
 & 10$^{-12}$ & 12.13 & 11.12 & 12.32 & 11.28 \\
 & 10$^{-13}$ & 13.05 & 11.83 & 13.07 & 11.84 \\
 & 10$^{-14}$ & 13.1 & 11.86 & 13.1 & 11.86 \\
\hline
\multirow{ 6}{*}{1000 $\times$ Solar} & 10$^{-9}$ & 1.02 & \textbf{1.14} & 1.03 & 1.35 \\
 & 10$^{-10}$ & 1.07 & 1.3 & 1.12 & 1.58 \\
 & 10$^{-11}$ & 3.33 & 3.78 & 3.52 & 4.02 \\
 & 10$^{-12}$ & 5.97 & 6.02 & 6.11 & 6.14 \\
 & 10$^{-13}$ & 6.77 & 6.65 & 6.79 & 6.67 \\
 & 10$^{-14}$ & 6.83 & 6.69 & 6.83 & 6.69 \\
\hline
\multirow{ 6}{*}{Steam atmosphere} & 10$^{-9}$ & 1.01 & \textbf{1.1} & 1.0 & 1.39 \\
 & 10$^{-10}$ & 1.15 & 1.67 & 1.09 & 1.3 \\
 & 10$^{-11}$ & 1.23 & 1.59 & 1.26 & 1.81 \\
 & 10$^{-12}$ & 3.01 & 3.68 & 3.29 & 3.97 \\
 & 10$^{-13}$ & 4.69 & 5.2 & 4.72 & 5.23 \\
 & 10$^{-14}$ & 4.78 & 5.28 & 4.78 & 5.28 \\
\hline
Flat line & \nodata &  0.95 & 1.40 & \nodata & \nodata  \\
\enddata
\tablenotetext{*}{Bolded cases represent the pure scattering haze models with the lowest $\chi_r^2$, which are plotted in Figure \ref{fig:nircam}. }
\end{deluxetable*}

The transmission spectrum from the near to the mid-IR is flat and featureless, and thus we are unable to constrain specific atmospheric compositions and detect specific molecular species. However, the flatness of the spectrum allows us to place limits on the atmospheric mean molecular weight and haze production rate by determining which models have sufficiently small spectral features. Our data-model comparisons suggest atmospheric mean molecular weights $\geq$15 g mol$^{-1}$ (i.e. the best fit models are the 1000 $\times$ solar metallicity and steam atmosphere cases) and haze production rates $\geq$10$^{-10}$ g cm$^{-2}$ s$^{-1}$ (Figures \ref{fig:spec}--\ref{fig:rcs}). The extension of the flat transmission spectrum into the mid-infrared disfavors lower mean molecular weights (e.g. lower metallicities), which were allowed by previous observations in the NIR for comparatively high haze production rates (Table \ref{table:rcs}). The high metallicities inferred here are consistent with core accretion population synthesis models that take into account envelope pollution by accreted planetesimals \citep{fortney2013}. We do not rule out non-solar ratio atmospheric compositions like the steam atmosphere scenario. While a flat line transmission spectrum results in a low $\chi_r^2$ of 1.4 (with only 1 degree of freedom), some of our models possess just as low values, if not lower (Table \ref{table:rcs}), though the differences in $\chi_r^2$ between these models are small.   

Some of the differences in goodness of fit between the best-fit models may be due to an unknown offset between the MIRI and the other data sets arising from differences in orbital and stellar parameters, limb-darkening coefficients, and stellar variability. As the slope of the spectrum from the NIR to the MIR is crucial in constraining our models, we tested for the effect of an offset by adding a free parameter that shifts the MIRI transit depths up and down. While the best fit cases had positive offsets of 50-100 ppm (i.e. increasing the transit depths at the MIRI wavelengths), the resulting $\chi_r^2$ and trend thereof with atmospheric composition and haze production rate were not qualitatively different from our nominal results.


We do not find any significant differences between the $\chi_r^2$'s of the purely scattering and Titan tholin hazes (Figure \ref{fig:rcs}). While the latter possess spectral features (Figures \ref{fig:opticalprops} and \ref{fig:specthol}), these features are fairly broad and low amplitude, particularly in the high mean molecular weight, high haze production rate cases, and thus do not impact the fits significantly. As the single scattering albedo of the haze inferred from the GJ 1214 b phase curve lies between those of a purely scattering haze and the Titan tholin haze \citep{kempton2023}, our results suggest that uncertainties regarding the actual optical properties of GJ 1214 b's hazes do not impact our inferences of the atmospheric composition and haze production rate from the observed transmission spectrum.

\section{Discussion}\label{sec:discussion}

Our results reinforce the picture that GJ 1214 b possesses a hazy atmosphere that is highly enriched in elements heavier than hydrogen and helium. In particular, the flatness of the transmission spectrum from the NIR to the MIR suggests an atmospheric mean molecular weight $\geq$15 g mol$^{-1}$, indicating metallicities $\geq$1000~$\times$~solar or the existence of a secondary atmosphere composed primarily of gases at least as heavy as water vapor. Simultaneously, the column haze production rate needs to be near the upper end of those estimated from photochemical models \citep{kawashima2019b,lavvas2019}, implying high haze formation efficiencies and/or the inclusion of a wide range of trace gases as haze precursors such as \ce{CO}, \ce{CO2}, and sulfur species in addition to hydrocarbons and nitriles \citep{horst2018b,he2020}. Our results, however, depends on several model assumptions regarding atmospheric composition and aerosol microphysics, which we discuss below.

\subsection{Sensitivity of Results to Background Atmosphere}\label{sec:atmsens}

We do not include haze feedback on the atmospheric thermal structure and composition. As such, we cannot confirm whether our best-fitting haze models produce a Bond albedo of 0.6 that is consistent with our initial assumptions and observations, nor can we ascertain whether our haze models can reproduce the thermal emission data from \citet{kempton2023}. A self-consistent treatment of haze feedback is needed, which we will defer to a future study. Instead, here we will evaluate how temperature errors associated with our lack of haze feedback contribute to errors in our model transmission spectra. Since spectral features in transmission are most sensitive to the scale height, which scales only linearly with temperature (versus to the fourth power for emission), we expect temperature errors to impact transmission spectroscopy minimally. We can estimate the effect of temperature errors on our results by considering that the transit depth difference $\Delta D$ associated with a single scale height is \citep{stevenson2016},


\begin{equation}
\label{eq:deltad}
\Delta D = \frac{2HR_p}{R_s^2}
\end{equation}
\noindent where $R_s$ is the stellar radius. Given the appropriate values in Table \ref{table:parameters} and $H$ $\sim$ 18 km from ${\S}$\ref{sec:hazeprop} for the 1000 $\times$ solar metallicity case, we get $\Delta D$ $\sim$ 30 ppm for one scale height, similar to the uncertainties on the HST WFC3 data \citep{kreidberg2014}. To evaluate the impact of our haze models having lower Bond albedo than the assumed value of 0.6, we note that the zero Bond albedo temperature of the planet $\sim$600 K \citep{cloutier2021}, 100 K higher than the photospheres of our model atmospheres (Figure \ref{fig:tpkzz}); this leads to an associated transit depth error of only $\sim(600-500)/500\times30~{\rm ppm}=6~{\rm ppm}$, which would not be detectable given current data quality. In other words, our conclusions here should not be affected if our haze models possessed lower Bond albedo than the assumed value of 0.6. While a thermal inversion of $>$100 K due to haze absorption in the upper atmosphere is certainly possible, the temperature increase would have to be several hundred K to be detectable in transmission, which may not be physical. Conversely, if our haze models possessed Bond albedo $>$ 0.6, then we would be overestimating their atmospheric temperature and therefore their scale height, and lower atmospheric mean molecular weight atmospheres would be allowed by the data once haze feedback is taken into account. However, such high albedo models would not be permitted by the thermal emission observations \citep{kempton2023} and as such they are not particularly relevant. Regardless of these potential temperature errors, the composition of the atmosphere is likely to be insensitive to haze feedback, as the temperature in the observable part of the atmosphere is solidly on the \ce{CH4} side of the \ce{CH4}/\ce{CO} transition (Figure \ref{fig:mrvis}), such that changing the local temperature by $\sim$100 K should not significantly alter the abundances of the main trace gases. In summary, even though haze feedback is likely vital for controlling the Bond albedo and thermal emission of GJ 1214 b, its transmission spectrum should not be strongly affected.



While our work focused on photochemical hazes, which are the result of disequilibrium chemical processes, the background gas composition of our model atmospheres -- and thus the primary spectral features in transmission -- were computed assuming thermochemical equilibrium. Disequilibrium chemical processes such as photochemistry and gas transport by mixing and advection are likely active on GJ 1214 b, as has been predicted by previous modeling works \citep{millerricci2012,morley2013,hu2014,morley2015,kawashima2018,kawashima2019b,lavvas2019}. These studies focused on hydrocarbon, oxygen, and nitrogen chemistry for various atmospheric metallicities and found significant production of higher order hydrocarbons and nitriles, though at very high metallicities ($\sim$1000 $\times$ solar) the high O/H ratio leads to a decrease in reducing species. Future modeling studies that include sulfur photochemistry \citep{tsai2022arXiv} that could enhance haze production \citep{he2020} should prove informative. However, it may be difficult to detect disequilibrium chemical species through transmission spectroscopy alone due to the small atmospheric scale height and large haze opacity of GJ 1214 b suggested by our study.


We examined a steam atmosphere scenario to evaluate the impact of non-solar ratio atmospheric compositions on the transmission spectrum. However, such an atmosphere is unlikely to reflect the actual atmospheric state of GJ 1214 b. For example, it would be difficult for a pure steam atmosphere to host hazes, unless contaminants like carbon, nitrogen, and sulfur were also present to form hydrocarbon, nitrile, and/or sulfur/sulfuric acid aerosols. The recent detection of \ce{He} escape from GJ 1214 b \citep{orellmiquel2022} would also rule out a ``secondary'' atmospheric composition, but the detection is controversial \citep{kasper2020,spake2022}.

\subsection{Sensitivity of Results to Haze Microphysics}\label{sec:hazesens}

In our study we considered spherical haze particles, but porous fractal aggregates are another form that haze particles can take \citep{west1991evidence}. Such ``fluffy'' particles can possess larger cross sections with which to block and scatter light, while staying aloft at lower pressures due to their lower densities compared to solid spherical particles. As a result, they can lead to flatter transmission spectra for a given haze production rate and background atmosphere \citep{adams2019}. However, the degree of porosity or aggregate fractal dimension is important in determining the particle extinction cross section and how it varies with wavelength -- at fractal dimensions $\sim$2, similar to those in Titan's atmosphere, the wavelength dependence of aggregates is determined by the (much smaller) monomer size rather than the total aggregate size, and thus a spectral slope is introduced \citep{lavvas2019,ohno2020agg}. There is thus a trade-off between higher porosity and increased wavelength dependence and it is therefore not obvious how introducing porous particles to our models would affect our constraints on GJ 1214 b's atmosphere. Future work is needed to determine whether including porous particles could reduce the haze production rate needed to explain GJ 1214 b's full transmission spectrum. 

Another assumption we made is that haze particle coagulation proceeds with perfect efficiency (i.e. a sticking efficiency of 1). A reduced sticking efficiency is possible if the haze particles are charged due to interactions with ions in the atmosphere; like-charged haze particles would repeal each other, resulting in less efficient coagulation \citep{lavvas2010}. The impact of such processes on our results would be a decrease in the mean particle size and an increase in haze number density, effectively increasing haze opacity overall but introducing more of a spectral slope due to the decrease in the ratio of particle radius to wavelength. Consequently, we would see reduced molecular feature strengths, as the haze becomes optically thick at lower pressures, while the spectral slope may become steeper; the former would increase the goodness of fit of the model to data while the latter would decrease the goodness of fit, and as such the net impact on $\chi_r^2$ is uncertain and would likely depend on the exact sticking efficiency. 

\subsection{The Role of Clouds}\label{sec:clouds}

\begin{figure*}[hbt!]
\centering
\includegraphics[width=0.8 \textwidth]{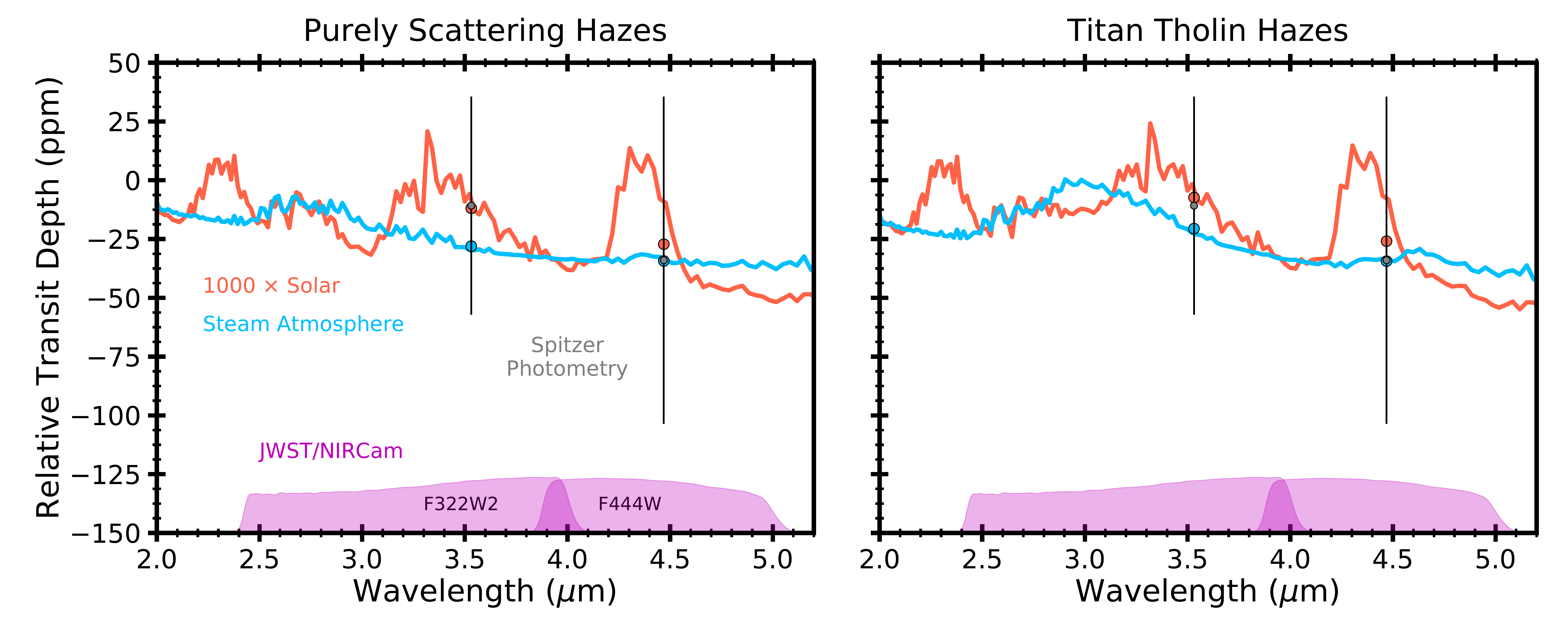}
\caption{Model transmission spectra for the two lowest $\chi_r^2$ models with purely scattering hazes (see Table \ref{table:rcs}; left) and corresponding Titan tholin hazes (right) in the 2 to 5 $\mu$m range compared to Spitzer photometry \citep{fraine2013} and the profiles of the JWST NIRCam filters F322W2 and F444W, which will be used to observe the transmission spectrum of GJ 1214b under JWST Cycle 1 GTO Program 1185 \citep{greene2017}.}
\label{fig:nircam}
\end{figure*}

We chose to focus on hazes in this work, but condensate clouds have also been invoked to explain the flat transmission spectra \citep{morley2013,morley2015,gao2018b,charnay2015b,ohno2018,ohno2020agg,christie2022}. However, as clouds form through the upwelling of condensate vapor from depth, as opposed to hazes that form naturally at low pressures near the UV photosphere, a greater amount of vertical transport is needed to bring the cloud particles up to the altitudes required. For example, \citet{gao2018b} needed a $K_{zz}$ of 10$^{10}$ cm$^2$ s$^{-1}$ and high atmospheric metallicity (1000 $\times$ solar) to match the NIR data with a microphysical model of KCl clouds, even though GCMs predict much lower $K_{zz}$ values \citep{charnay2015a}. \citet{ohno2018} considered the GCM-derived $K_{zz}$ in their microphysical model of KCl clouds and found that they could not reproduce the NIR data even with a high mean molecular weight steam atmosphere. The inclusion of porous aggregate cloud particles in their follow-up work \citep{ohno2020agg} allowed their models to match the NIR data with an atmospheric metallicity of 1000 $\times$ solar, but it is unknown whether their data-model agreement extends to MIRI wavelengths. 

Modeling efforts that used the \texttt{eddysed} framework \citep{ackerman2001,rooney2022} and GCM-derived $K_{zz}$ needed very low (0.1-0.01) values for the sedimentation efficiency parameter, $f_{\rm sed}$, to produce the cloud vertical extents required by the data \citep{morley2015,christie2022}. \citet{charnay2015b} included KCl and ZnS clouds in their GCM study and was able to match the NIR data with an atmospheric metallicity of 100 $\times$ solar, but their model spectra possessed large spectral variations in the mid-infrared that would be rejected by the MIRI data. In all, matching the NIR and MIR data with clouds alone would likely be much more difficult than doing so with hazes under reasonable model assumptions. 

An interesting scenario that could be explored in future studies is the interaction of clouds and hazes. Organic hazes are common cloud condensation nuclei on Earth \citep{sun2006} and water cloud formation typically removes these haze particles from our atmosphere \citep[e.g.][]{oduber2021}. \citet{yu2021} studied the surface energies of different laboratory exoplanet haze analogues and found that hazes formed at high atmospheric metallicity and at the temperature of GJ 1214 b may be especially difficult for KCl to nucleate on due to their low surface energies, suggesting minimal haze removal by KCl cloud formation. However, if the surface energy of the actual hazes in the atmosphere of GJ 1214 b were higher, and sedimentation were the primary cloud transport process instead of mixing, then haze removal via KCl cloud formation may be efficient, reducing the haze opacity such that increases in the haze production rate and/or atmospheric mean molecular weight would be needed to match the flat transmission spectrum.

\subsection{Implications for Internal Structure}\label{sec:internal}


The constraint on the atmospheric mean molecular weight resulting from our study may have consequences for inferences of the bulk composition and internal structure of GJ 1214 b. As an example, here we assess whether a water world composition, as would be implied by a steam atmosphere, is realistic. To do this, we use the internal structure model described in \citet{nixon2021}, which has previously been applied to a number of sub-Neptunes \citep{luque2021,luque2022_g940b}, to estimate the bulk mass fraction of \ce{H2O} that would be required to explain the observed mass and radius of the planet. In this case, the model consists of a water envelope above an Earth-like core consisting of 1/3 \ce{Fe} and 2/3 \ce{MgSiO3} by mass. The temperature profile in the interior is computed by extending the TP profile of the steam atmosphere model (Figure \ref{fig:tpkzz}) into the interior, assuming an adiabatic temperature profile for pressures $>$100 bar. We found that, assuming a nominal photospheric pressure of 10 mbar, a water mass fraction of 73.7\% is needed to reproduce the planetary radius of 2.628 R$_{\oplus}$ (Table \ref{table:parameters}). The mass and radius of the planet can therefore be explained without invoking a H/He-dominated atmosphere, albeit at a somewhat higher mass fraction of water than has been suggested for other water worlds \citep[$\sim$50\%;][]{luque2022}. Future work will examine the full range of compositions that are possible for this planet in light of the new insights into its atmosphere provided in \citet{kempton2023} and the present study.

\subsection{Outlook for Future Observations}\label{sec:futureobs}

Additional transmission spectra of GJ 1214 b may soon be obtained within the F322W2 and F444W filters of JWST NIRCam through GTO program 1185 \citep{greene2017}. Based on our best fitting models, we predict feature amplitudes of a few tens of ppm within those filters, with about the same magnitude of difference between the models (Figure \ref{fig:nircam}). While tens of ppm is likely above the noise floor of NIRCam \citep{ahrer2022,schlawin2023}, the dimness of the host star GJ 1214 (J = 9.75) may require multiple transits and/or significant spectral binning to recover the molecular features. Some differences also exist between the purely scattering haze and Titan tholin haze cases due to the latter exhibiting broad haze spectral features at 3 and 4.5 $\mu$m, but these differences are on the order of $\sim$10 ppm.

The inferred high albedo of GJ 1214 b suggests that reflected light observations may be a potential alternative avenue for atmospheric characterization. The planet-star contrast near secondary eclipse (planetary phase angle = 0$^{\circ}$), $C_0$, can be estimated using \citep{cahoy2010}

\begin{equation}
\label{eq:c0}
C_0 = \frac23 A_g(\lambda)\left (\frac{R_p}{a} \right )^2
\end{equation}

\noindent where $A_g(\lambda)$ is the wavelength-dependent geometric albedo and $a$ is the planet semi-major axis. Eq. \ref{eq:c0} assumes that the planet reflects light isotropically. Using the values from Table \ref{table:parameters}, we find $C_0
\sim(41~{\rm ppm}) \times A_g(\lambda)$ for GJ 1214 b. For reference, $A_g(\lambda)=2/3$ for a perfectly reflecting Lambert sphere, 3/4 for a purely Rayleigh scattering atmosphere, and could reach $>$1 for anisotropically scattering atmospheres, though that would invalidate the assumption inherent in Eq. \ref{eq:c0}  \citep{cahoy2010}. While precision of a few tens of ppm appears to be above the noise floor of JWST's near-IR instruments \citep{rustamkulov2022,schlawin2023,coulombe2023arXiv}, as with the NIRCam measurements, multiple transits and spectral binning are likely needed to reach the necessary SNR for detecting reflected light. However, without a self-consistent computation of the thermal emission of the planet (${\S}$\ref{sec:atmsens}), the wavelengths at which reflected light dominates over thermal emission is uncertain. As such, we leave a more detailed, self-consistent calculation of the reflected light and thermal emission spectra of our haze models, as well as the resultant Bond albedo to a future study.  



As a complementary technique, ground-based high-resolution transmission spectroscopy may be able to see the cores of molecular spectral lines that stick up above the haze layer of GJ 1214 b, allowing for characterization of its atmospheric composition \citep{hood2020,gandhi2020,lafarga2023arXiv}. In particular, \citet{hood2020} considered a 50 $\times$ solar metallicity background atmosphere for GJ 1214 b and a haze that became opaque at 10 $\mu$bar, similar to our highest haze production rate cases. They also found that their modeled observations are most sensitive to the atmosphere at 1 $\mu$bar, where hazes are optically thin in our models. However, the much lower metallicity of their models mean that the scale height of our model atmospheres are only $\sim1/5$ of theirs. As such, even though they found that ground-based observations on both current and future facilities should be able to detect certain molecules in GJ 1214 b's atmosphere on reasonable timescales, the impact of the decreased scale heights we have obtained requires further evaluation.  

\section{Conclusions}\label{sec:conclusions}


Even though the transmission spectrum of GJ 1214 b remains flat and featureless, the addition of mid-infrared data has allowed us to reject all but the highest metallicity/atmospheric mean molecular weight ($\geq$15 g mol$^{-1}$) and haze production rate ($\geq$10$^{-10}$ g cm$^{-2}$ s$^{-1}$) models, pointing to an atmosphere dominated by species heavier than hydrogen and helium and efficient high altitude aerosol formation. Further characterization of GJ 1214 b's atmosphere through transmission spectroscopy will likely require JWST observations of multiple transits, while the same will be true of secondary eclipses to extract additional information from dayside emission spectra \citep{kempton2023}. Alternatively, reflected light from GJ 1214 b may be detectable due to the presence of its high albedo aerosol layer, while ground-based high resolution transmission spectroscopy could serve as a complementary technique to characterize GJ 1214 b's atmospheric composition. Future self-consistent modeling that includes haze radiative feedback on the atmospheric thermal structure will be needed to assess the viability of observational strategies beyond transmission spectroscopy. Taken together, the JWST MIRI LRS observations of GJ 1214 b demonstrate the importance of aerosols and high metallicities for interpreting future JWST observations of warm sub-Neptune exoplanets, necessitating careful observational planning in our quest to understand the most abundant type of planet in the Galaxy.

\acknowledgments


 We thank H. Zhang, W. Z. Gao, and S. Z. Gao for their loving support during the writing of this paper. This work is based on observations made with the NASA/ESA/CSA JWST. The data were obtained from the Mikulski Archive for Space Telescopes at the Space Telescope Science Institute, which is operated by the Association of Universities for Research in Astronomy, Inc., under NASA contract NAS 5-03127 for JWST. These observations are associated with program \#1803. Support for this program was provided by NASA through a grant from the Space Telescope Science Institute. 




\vspace{5mm}
\facilities{JWST(MIRI), HST(WFC3), Spitzer}


\bibliography{references}

\begin{thebibliography}{}
\expandafter\ifx\csname natexlab\endcsname\relax\def\natexlab#1{#1}\fi
\providecommand{\url}[1]{\href{#1}{#1}}
\providecommand{\dodoi}[1]{doi:~\href{http://doi.org/#1}{\nolinkurl{#1}}}
\providecommand{\doeprint}[1]{\href{http://ascl.net/#1}{\nolinkurl{http://ascl.net/#1}}}
\providecommand{\doarXiv}[1]{\href{https://arxiv.org/abs/#1}{\nolinkurl{https://arxiv.org/abs/#1}}}

\bibitem[{Ackerman \& Marley(2001)}]{ackerman2001}
Ackerman, A.~S., \& Marley, M.~S. 2001, \apj, 556, 872, \dodoi{10.1086/321540}

\bibitem[{{Ackerman} {et~al.}(1995){Ackerman}, {Toon}, \&
  {Hobbs}}]{ackerman1995}
{Ackerman}, A.~S., {Toon}, O.~B., \& {Hobbs}, P.~V. 1995, \jgr, 100, 7121,
  \dodoi{10.1029/95JD00026}

\bibitem[{{Adams} {et~al.}(2019){Adams}, {Gao}, {de Pater}, \&
  {Morley}}]{adams2019}
{Adams}, D., {Gao}, P., {de Pater}, I., \& {Morley}, C.~V. 2019, \apj, 874, 61,
  \dodoi{10.3847/1538-4357/ab074c}

\bibitem[{{Ahrer} {et~al.}(2022){Ahrer}, {Stevenson}, {Mansfield}, {Moran},
  {Brande}, {Morello}, {Murray}, {Nikolov}, {Petit dit de la Roche},
  {Schlawin}, {Wheatley}, {Zieba}, {Batalha}, {Damiano}, {Goyal}, {Lendl},
  {Lothringer}, {Mukherjee}, {Ohno}, {Batalha}, {Battley}, {Bean}, {Beatty},
  {Benneke}, {Berta-Thompson}, {Carter}, {Cubillos}, {Daylan}, {Espinoza},
  {Gao}, {Gibson}, {Gill}, {Harrington}, {Hu}, {Kreidberg}, {Lewis}, {Line},
  {L{\'o}pez-Morales}, {Parmentier}, {Powell}, {Sing}, {Tsai}, {Wakeford},
  {Welbanks}, {Alam}, {Alderson}, {Allen}, {Anderson}, {Barstow}, {Bayliss},
  {Bell}, {Blecic}, {Bryant}, {Burleigh}, {Carone}, {Casewell}, {Changeat},
  {Chubb}, {Crossfield}, {Crouzet}, {Decin}, {D{\'e}sert}, {Feinstein},
  {Flagg}, {Fortney}, {Gizis}, {Heng}, {Iro}, {Kempton}, {Kendrew}, {Kirk},
  {Knutson}, {Komacek}, {Lagage}, {Leconte}, {Lustig-Yaeger}, {MacDonald},
  {Mancini}, {May}, {Mayne}, {Miguel}, {Mikal-Evans}, {Molaverdikhani},
  {Palle}, {Piaulet}, {Rackham}, {Redfield}, {Rogers}, {Roy}, {Rustamkulov},
  {Shkolnik}, {Sotzen}, {Taylor}, {Tremblin}, {Tucker}, {Turner}, {de
  Val-Borro}, {Venot}, \& {Zhang}}]{ahrer2022}
{Ahrer}, E.-M., {Stevenson}, K.~B., {Mansfield}, M., {et~al.} 2022, arXiv
  e-prints, arXiv:2211.10489, \dodoi{10.48550/arXiv.2211.10489}

\bibitem[{{Asplund} {et~al.}(2009){Asplund}, {Grevesse}, {Sauval}, \&
  {Scott}}]{asplund2009}
{Asplund}, M., {Grevesse}, N., {Sauval}, A.~J., \& {Scott}, P. 2009, \araa, 47,
  481, \dodoi{10.1146/annurev.astro.46.060407.145222}

\bibitem[{{Barklem} \& {Collet}(2016)}]{barklem16}
{Barklem}, P.~S., \& {Collet}, R. 2016, \aap, 588, A96,
  \dodoi{10.1051/0004-6361/201526961}

\bibitem[{{Batalha} {et~al.}(2017){Batalha}, {Mandell}, {Pontoppidan},
  {Stevenson}, {Lewis}, {Kalirai}, {Earl}, {Greene}, {Albert}, \&
  {Nielsen}}]{batalha2017}
{Batalha}, N.~E., {Mandell}, A., {Pontoppidan}, K., {et~al.} 2017, \pasp, 129,
  064501, \dodoi{10.1088/1538-3873/aa65b0}

\bibitem[{{Bean} {et~al.}(2010){Bean}, {Miller-Ricci Kempton}, \&
  {Homeier}}]{bean2010}
{Bean}, J.~L., {Miller-Ricci Kempton}, E., \& {Homeier}, D. 2010, \nat, 468,
  669, \dodoi{10.1038/nature09596}

\bibitem[{{Bean} {et~al.}(2011){Bean}, {D{\'e}sert}, {Kabath}, {Stalder},
  {Seager}, {Miller-Ricci Kempton}, {Berta}, {Homeier}, {Walsh}, \&
  {Seifahrt}}]{bean2011}
{Bean}, J.~L., {D{\'e}sert}, J.-M., {Kabath}, P., {et~al.} 2011, \apj, 743, 92,
  \dodoi{10.1088/0004-637X/743/1/92}

\bibitem[{{Berta} {et~al.}(2012){Berta}, {Charbonneau}, {D{\'e}sert},
  {Miller-Ricci Kempton}, {McCullough}, {Burke}, {Fortney}, {Irwin}, {Nutzman},
  \& {Homeier}}]{berta2012}
{Berta}, Z.~K., {Charbonneau}, D., {D{\'e}sert}, J.-M., {et~al.} 2012, \apj,
  747, 35, \dodoi{10.1088/0004-637X/747/1/35}

\bibitem[{Bohren \& Huffman(2008)}]{bohren2008}
Bohren, C., \& Huffman, D. 2008, Absorption and Scattering of Light by Small
  Particles, Wiley Science Series (Wiley).
\newblock \url{https://books.google.com/books?id=ib3EMXXIRXUC}

\bibitem[{{Brown}(2001)}]{Brown2001}
{Brown}, T.~M. 2001, \apj, 553, 1006, \dodoi{10.1086/320950}

\bibitem[{Cahoy {et~al.}(2010)Cahoy, Marley, \& Fortney}]{cahoy2010}
Cahoy, K.~L., Marley, M.~S., \& Fortney, J.~J. 2010, \apj, 724, 189,
  \dodoi{10.1088/0004-637X/724/1/189}

\bibitem[{{Charbonneau} {et~al.}(2009){Charbonneau}, {Berta}, {Irwin}, {Burke},
  {Nutzman}, {Buchhave}, {Lovis}, {Bonfils}, {Latham}, {Udry}, {Murray-Clay},
  {Holman}, {Falco}, {Winn}, {Queloz}, {Pepe}, {Mayor}, {Delfosse}, \&
  {Forveille}}]{charbonneau2009}
{Charbonneau}, D., {Berta}, Z.~K., {Irwin}, J., {et~al.} 2009, \nat, 462, 891

\bibitem[{{Charnay} {et~al.}(2015{\natexlab{a}}){Charnay}, {Meadows}, \&
  {Leconte}}]{charnay2015a}
{Charnay}, B., {Meadows}, V., \& {Leconte}, J. 2015{\natexlab{a}}, \apj, 813,
  15

\bibitem[{{Charnay} {et~al.}(2015{\natexlab{b}}){Charnay}, {Meadows}, {Misra},
  {Leconte}, \& {Arney}}]{charnay2015b}
{Charnay}, B., {Meadows}, V., {Misra}, A., {Leconte}, J., \& {Arney}, G.
  2015{\natexlab{b}}, \apjl, 813, L1

\bibitem[{{Checlair} {et~al.}(2016){Checlair}, {McKay}, \&
  {Imanaka}}]{checlair2016}
{Checlair}, J., {McKay}, C.~P., \& {Imanaka}, H. 2016, \planss, 129, 1,
  \dodoi{10.1016/j.pss.2016.03.012}

\bibitem[{{Christie} {et~al.}(2022){Christie}, {Mayne}, {Gillard}, {Manners},
  {H{\'e}brard}, {Lines}, \& {Kohary}}]{christie2022}
{Christie}, D.~A., {Mayne}, N.~J., {Gillard}, R.~M., {et~al.} 2022, \mnras,
  517, 1407, \dodoi{10.1093/mnras/stac2763}

\bibitem[{{Cloutier} {et~al.}(2021){Cloutier}, {Charbonneau}, {Deming},
  {Bonfils}, \& {Astudillo-Defru}}]{cloutier2021}
{Cloutier}, R., {Charbonneau}, D., {Deming}, D., {Bonfils}, X., \&
  {Astudillo-Defru}, N. 2021, \aj, 162, 174, \dodoi{10.3847/1538-3881/ac1584}

\bibitem[{{Col{\'o}n} \& {Gaidos}(2013)}]{colon2013}
{Col{\'o}n}, K.~D., \& {Gaidos}, E. 2013, \apj, 776, 49,
  \dodoi{10.1088/0004-637X/776/1/49}

\bibitem[{{Corrales} {et~al.}(2023){Corrales}, {Gavilan}, {Teal}, \&
  {Kempton}}]{corrales2023}
{Corrales}, L., {Gavilan}, L., {Teal}, D.~J., \& {Kempton}, E. M.~R. 2023,
  \apjl, 943, L26, \dodoi{10.3847/2041-8213/acaf86}

\bibitem[{{Coulombe} {et~al.}(2023){Coulombe}, {Benneke}, {Challener},
  {Piette}, {Wiser}, {Mansfield}, {MacDonald}, {Beltz}, {Feinstein}, {Radica},
  {Savel}, {Dos Santos}, {Bean}, {Parmentier}, {Wong}, {Rauscher}, {Komacek},
  {Kempton}, {Tan}, {Hammond}, {Lewis}, {Line}, {Lee}, {Shivkumar},
  {Crossfield}, {Nixon}, {Rackham}, {Wakeford}, {Welbanks}, {Zhang}, {Batalha},
  {Berta-Thompson}, {Changeat}, {D{\'e}sert}, {Espinoza}, {Goyal},
  {Harrington}, {Knutson}, {Kreidberg}, {L{\'o}pez-Morales}, {Shporer}, {Sing},
  {Stevenson}, {Aggarwal}, {Ahrer}, {Alam}, {Bell}, {Blecic}, {Caceres},
  {Carter}, {Casewell}, {Crouzet}, {Cubillos}, {Decin}, {Fortney}, {Gibson},
  {Heng}, {Henning}, {Iro}, {Kendrew}, {Lagage}, {Leconte}, {Lendl},
  {Lothringer}, {Mancini}, {Mikal-Evans}, {Molaverdikhani}, {Nikolov}, {Ohno},
  {Palle}, {Piaulet}, {Redfield}, {Roy}, {Tsai}, {Venot}, \&
  {Wheatley}}]{coulombe2023arXiv}
{Coulombe}, L.-P., {Benneke}, B., {Challener}, R., {et~al.} 2023, arXiv
  e-prints, arXiv:2301.08192, \dodoi{10.48550/arXiv.2301.08192}

\bibitem[{{Crossfield} {et~al.}(2011){Crossfield}, {Barman}, \&
  {Hansen}}]{crossfield2011}
{Crossfield}, I.~J.~M., {Barman}, T., \& {Hansen}, B. M.~S. 2011, \apj, 736,
  132, \dodoi{10.1088/0004-637X/736/2/132}

\bibitem[{Davidson(1993)}]{davidson1993}
Davidson, T. 1993, A Simple and Accurate Method for Calculating Viscosity of
  Gaseous Mixtures, Report of investigations (U.S. Department of the Interior,
  Bureau of Mines).
\newblock \url{https://books.google.com/books?id=N3BwoKcR4xYC}

\bibitem[{{de Mooij} {et~al.}(2012){de Mooij}, {Brogi}, {de Kok},
  {Koppenhoefer}, {Nefs}, {Snellen}, {Greiner}, {Hanse}, {Heinsbroek}, {Lee},
  \& {van der Werf}}]{demooij2012}
{de Mooij}, E.~J.~W., {Brogi}, M., {de Kok}, R.~J., {et~al.} 2012, \aap, 538,
  A46, \dodoi{10.1051/0004-6361/201117205}

\bibitem[{{D{\'e}sert} {et~al.}(2011){D{\'e}sert}, {Bean}, {Miller-Ricci
  Kempton}, {Berta}, {Charbonneau}, {Irwin}, {Fortney}, {Burke}, \&
  {Nutzman}}]{desert2011}
{D{\'e}sert}, J.-M., {Bean}, J., {Miller-Ricci Kempton}, E., {et~al.} 2011,
  \apjl, 731, L40, \dodoi{10.1088/2041-8205/731/2/L40}

\bibitem[{{Drummond} {et~al.}(2018){Drummond}, {Mayne}, {Baraffe}, {Tremblin},
  {Manners}, {Amundsen}, {Goyal}, \& {Acreman}}]{drummond2018met}
{Drummond}, B., {Mayne}, N.~J., {Baraffe}, I., {et~al.} 2018, \aap, 612, A105,
  \dodoi{10.1051/0004-6361/201732010}

\bibitem[{Fortney {et~al.}(2005)Fortney, Marley, Lodders, Saumon, \&
  Freedman}]{fortney2005}
Fortney, J.~J., Marley, M.~S., Lodders, K., Saumon, D., \& Freedman, R. 2005,
  \apj, 627, L69, \dodoi{10.1086/431952}

\bibitem[{{Fortney} {et~al.}(2013){Fortney}, {Mordasini}, {Nettelmann},
  {Kempton}, {Greene}, \& {Zahnle}}]{fortney2013}
{Fortney}, J.~J., {Mordasini}, C., {Nettelmann}, N., {et~al.} 2013, \apj, 775,
  80, \dodoi{10.1088/0004-637X/775/1/80}

\bibitem[{{Fortney} {et~al.}(2003){Fortney}, {Sudarsky}, {Hubeny}, {Cooper},
  {Hubbard}, {Burrows}, \& {Lunine}}]{fortney2003}
{Fortney}, J.~J., {Sudarsky}, D., {Hubeny}, I., {et~al.} 2003, \apj, 589, 615,
  \dodoi{10.1086/374387}

\bibitem[{{Fraine} {et~al.}(2013){Fraine}, {Deming}, {Gillon}, {Jehin},
  {Demory}, {Benneke}, {Seager}, {Lewis}, {Knutson}, \&
  {D{\'e}sert}}]{fraine2013}
{Fraine}, J.~D., {Deming}, D., {Gillon}, M., {et~al.} 2013, \apj, 765, 127,
  \dodoi{10.1088/0004-637X/765/2/127}

\bibitem[{{Freedman} {et~al.}(2014){Freedman}, {Lustig-Yaeger}, {Fortney},
  {Lupu}, {Marley}, \& {Lodders}}]{freedman2014}
{Freedman}, R.~S., {Lustig-Yaeger}, J., {Fortney}, J.~J., {et~al.} 2014, \apjs,
  214, 25, \dodoi{10.1088/0067-0049/214/2/25}

\bibitem[{{Gandhi} {et~al.}(2020){Gandhi}, {Brogi}, \& {Webb}}]{gandhi2020}
{Gandhi}, S., {Brogi}, M., \& {Webb}, R.~K. 2020, \mnras, 498, 194,
  \dodoi{10.1093/mnras/staa2424}

\bibitem[{{Gandhi} \& {Madhusudhan}(2017)}]{gandhi2017}
{Gandhi}, S., \& {Madhusudhan}, N. 2017, \mnras, 472, 2334,
  \dodoi{10.1093/mnras/stx1601}

\bibitem[{{Gao} \& {Benneke}(2018)}]{gao2018b}
{Gao}, P., \& {Benneke}, B. 2018, \apj, 863, 165,
  \dodoi{10.3847/1538-4357/aad461}

\bibitem[{{Gao} {et~al.}(2018){Gao}, {Marley}, \& {Ackerman}}]{gao2018a}
{Gao}, P., {Marley}, M.~S., \& {Ackerman}, A.~S. 2018, \apj, 855, 86,
  \dodoi{10.3847/1538-4357/aab0a1}

\bibitem[{{Gao} {et~al.}(2021){Gao}, {Wakeford}, {Moran}, \&
  {Parmentier}}]{gao2021rev}
{Gao}, P., {Wakeford}, H.~R., {Moran}, S.~E., \& {Parmentier}, V. 2021, Journal
  of Geophysical Research (Planets), 126, e06655, \dodoi{10.1029/2020JE006655}

\bibitem[{{Gao} \& {Zhang}(2020)}]{gao2020sp}
{Gao}, P., \& {Zhang}, X. 2020, \apj, 890, 93, \dodoi{10.3847/1538-4357/ab6a9b}

\bibitem[{{Gharib-Nezhad} {et~al.}(2021){Gharib-Nezhad}, {Iyer}, {Line},
  {Freedman}, {Marley}, \& {Batalha}}]{GharibNezhad2021}
{Gharib-Nezhad}, E., {Iyer}, A.~R., {Line}, M.~R., {et~al.} 2021, \apjs, 254,
  34, \dodoi{10.3847/1538-4365/abf504}

\bibitem[{{Gordon} {et~al.}(2017{\natexlab{a}}){Gordon}, {Rothman}, {Tan},
  {Kochanov}, \& {Hill}}]{HITRAN2016}
{Gordon}, I.~E., {Rothman}, L.~S., {Tan}, Y., {Kochanov}, R.~V., \& {Hill}, C.
  2017{\natexlab{a}}, in 72nd International Symposium on Molecular
  Spectroscopy, TJ08, \dodoi{10.15278/isms.2017.TJ08}

\bibitem[{{Gordon} {et~al.}(2017{\natexlab{b}}){Gordon}, {Rothman}, {Hill},
  {Kochanov}, {Tan}, {Bernath}, {Birk}, {Boudon}, {Campargue}, {Chance},
  {Drouin}, {Flaud}, {Gamache}, {Hodges}, {Jacquemart}, {Perevalov}, {Perrin},
  {Shine}, {Smith}, {Tennyson}, {Toon}, {Tran}, {Tyuterev}, {Barbe},
  {Cs{\'a}sz{\'a}r}, {Devi}, {Furtenbacher}, {Harrison}, {Hartmann}, {Jolly},
  {Johnson}, {Karman}, {Kleiner}, {Kyuberis}, {Loos}, {Lyulin}, {Massie},
  {Mikhailenko}, {Moazzen-Ahmadi}, {M{\"u}ller}, {Naumenko}, {Nikitin},
  {Polyansky}, {Rey}, {Rotger}, {Sharpe}, {Sung}, {Starikova}, {Tashkun},
  {Auwera}, {Wagner}, {Wilzewski}, {Wcis{\l}o}, {Yu}, \& {Zak}}]{gordon2017}
{Gordon}, I.~E., {Rothman}, L.~S., {Hill}, C., {et~al.} 2017{\natexlab{b}},
  \jqsrt, 203, 3, \dodoi{10.1016/j.jqsrt.2017.06.038}

\bibitem[{{Greene} {et~al.}(2017){Greene}, {Beatty}, {Rieke}, \&
  {Schlawin}}]{greene2017}
{Greene}, T.~P., {Beatty}, T.~G., {Rieke}, M.~J., \& {Schlawin}, E. 2017,
  {Transit Spectroscopy of Mature Planets}, JWST Proposal. Cycle 1, ID. \#1185

\bibitem[{{Harris} {et~al.}(2006){Harris}, {Tennyson}, {Kaminsky}, {Pavlenko},
  \& {Jones}}]{harris2006}
{Harris}, G.~J., {Tennyson}, J., {Kaminsky}, B.~M., {Pavlenko}, Y.~V., \&
  {Jones}, H.~R.~A. 2006, \mnras, 367, 400,
  \dodoi{10.1111/j.1365-2966.2005.09960.x}

\bibitem[{{He} {et~al.}(2020){He}, {H{\"o}rst}, {Lewis}, {Yu}, {Moses},
  {McGuiggan}, {Marley}, {Kempton}, {Moran}, {Morley}, \& {Vuitton}}]{he2020}
{He}, C., {H{\"o}rst}, S.~M., {Lewis}, N.~K., {et~al.} 2020, Nature Astronomy,
  \dodoi{10.1038/s41550-020-1072-9}

\bibitem[{{He} {et~al.}(2023){He}, {Radke}, {Moran}, {Horst}, {Lewis}, {Moses},
  {Marley}, {Batalha}, {Kempton}, {Morley}, {Valenti}, \&
  {Vuitton}}]{he2023arXiv}
{He}, C., {Radke}, M., {Moran}, S.~E., {et~al.} 2023, arXiv e-prints,
  arXiv:2301.02745, \dodoi{10.48550/arXiv.2301.02745}

\bibitem[{{Hood} {et~al.}(2020){Hood}, {Fortney}, {Line}, {Martin}, {Morley},
  {Birkby}, {Rustamkulov}, {Lupu}, \& {Freedman}}]{hood2020}
{Hood}, C.~E., {Fortney}, J.~J., {Line}, M.~R., {et~al.} 2020, \aj, 160, 198,
  \dodoi{10.3847/1538-3881/abb46b}

\bibitem[{{H{\"o}rst} {et~al.}(2018){H{\"o}rst}, {He}, {Lewis}, {Kempton},
  {Marley}, {Morley}, {Moses}, {Valenti}, \& {Vuitton}}]{horst2018b}
{H{\"o}rst}, S.~M., {He}, C., {Lewis}, N.~K., {et~al.} 2018, Nature Astronomy,
  2, 303, \dodoi{10.1038/s41550-018-0397-0}

\bibitem[{Hu \& Seager(2014)}]{hu2014}
Hu, R., \& Seager, S. 2014, \apj, 784, 63.
\newblock \url{http://stacks.iop.org/0004-637X/784/i=1/a=63}

\bibitem[{{Huang} {et~al.}(2014){Huang}, Gamache, Freedman, Schwenke, \&
  Lee}]{HUANG2014reliable}
{Huang}, X., Gamache, R.~R., Freedman, R.~S., Schwenke, D.~W., \& Lee, T.~J.
  2014, Journal of Quantitative Spectroscopy and Radiative Transfer, 147, 134 ,
  \dodoi{https://doi.org/10.1016/j.jqsrt.2014.05.015}

\bibitem[{Jacobson \& Turco(1994)}]{jacobson1994}
Jacobson, M.~Z., \& Turco, R.~P. 1994, Atmospheric Environment, 28, 1327

\bibitem[{{Kasper} {et~al.}(2020){Kasper}, {Bean}, {Oklop{\v{c}}i{\'c}},
  {Malsky}, {Kempton}, {D{\'e}sert}, {Rogers}, \& {Mansfield}}]{kasper2020}
{Kasper}, D., {Bean}, J.~L., {Oklop{\v{c}}i{\'c}}, A., {et~al.} 2020, \aj, 160,
  258, \dodoi{10.3847/1538-3881/abbee6}

\bibitem[{{Kataria} {et~al.}(2014){Kataria}, {Showman}, {Fortney}, {Marley}, \&
  {Freedman}}]{kataria2014}
{Kataria}, T., {Showman}, A.~P., {Fortney}, J.~J., {Marley}, M.~S., \&
  {Freedman}, R.~S. 2014, \apj, 785, 92, \dodoi{10.1088/0004-637X/785/2/92}

\bibitem[{{Kawashima} {et~al.}(2019){Kawashima}, {Hu}, \&
  {Ikoma}}]{kawashima2019a}
{Kawashima}, Y., {Hu}, R., \& {Ikoma}, M. 2019, \apjl, 876, L5,
  \dodoi{10.3847/2041-8213/ab16f6}

\bibitem[{{Kawashima} \& {Ikoma}(2018)}]{kawashima2018}
{Kawashima}, Y., \& {Ikoma}, M. 2018, \apj, 853, 7,
  \dodoi{10.3847/1538-4357/aaa0c5}

\bibitem[{{Kawashima} \& {Ikoma}(2019)}]{kawashima2019b}
---. 2019, \apj, 877, 109, \dodoi{10.3847/1538-4357/ab1b1d}

\bibitem[{{Kempton} {et~al.}(2023){Kempton}, {Zhang}, {Bean}, , E., {Piette},
  {Parmentier}, {Malsky}, {Roman}, \& {Rauscher}}]{kempton2023}
{Kempton}, E., {Zhang}, M., {Bean}, J.~L., {et~al.} 2023, submitted

\bibitem[{{Kendrew} {et~al.}(2015){Kendrew}, {Scheithauer}, {Bouchet},
  {Amiaux}, {Azzollini}, {Bouwman}, {Chen}, {Dubreuil}, {Fischer}, {Glasse},
  {Greene}, {Lagage}, {Lahuis}, {Ronayette}, {Wright}, \&
  {Wright}}]{kendrew2015}
{Kendrew}, S., {Scheithauer}, S., {Bouchet}, P., {et~al.} 2015, \pasp, 127,
  623, \dodoi{10.1086/682255}

\bibitem[{Khare {et~al.}(1984)Khare, Sagan, Arakawa, Suits, Callcott, \&
  Williams}]{khare1984}
Khare, B.~N., Sagan, C., Arakawa, K.~T., {et~al.} 1984, \icarus, 60, 127

\bibitem[{Kreidberg {et~al.}(2014)Kreidberg, Bean, D{\'{e}}sert, Benneke,
  Deming, Stevenson, Seager, Berta-Thompson, Seifahrt, \&
  Homeier}]{kreidberg2014}
Kreidberg, L., Bean, J.~L., D{\'{e}}sert, J.-M., {et~al.} 2014, \nat, 505, 69,
  \dodoi{10.1038/nature12888}

\bibitem[{{Lafarga} {et~al.}(2023){Lafarga}, {Brogi}, {Gandhi}, {Cegla},
  {Seidel}, {Doyle}, {Allart}, {Buchschacher}, {Lendl}, {Lovis}, \&
  {Sosnowska}}]{lafarga2023arXiv}
{Lafarga}, M., {Brogi}, M., {Gandhi}, S., {et~al.} 2023, arXiv e-prints,
  arXiv:2302.04794, \dodoi{10.48550/arXiv.2302.04794}

\bibitem[{{Lavvas} \& {Koskinen}(2017)}]{lavvas2017}
{Lavvas}, P., \& {Koskinen}, T. 2017, \apj, 847, 32,
  \dodoi{10.3847/1538-4357/aa88ce}

\bibitem[{{Lavvas} {et~al.}(2019){Lavvas}, {Koskinen}, {Steinrueck},
  {Garc{\'\i}a Mu{\~n}oz}, \& {Showman}}]{lavvas2019}
{Lavvas}, P., {Koskinen}, T., {Steinrueck}, M.~E., {Garc{\'\i}a Mu{\~n}oz}, A.,
  \& {Showman}, A.~P. 2019, \apj, 878, 118, \dodoi{10.3847/1538-4357/ab204e}

\bibitem[{Lavvas {et~al.}(2010)Lavvas, Yelle, \& Griffith}]{lavvas2010}
Lavvas, P., Yelle, R.~V., \& Griffith, C.~A. 2010, \icarus, 210, 832

\bibitem[{{Li} {et~al.}(2015){Li}, {Gordon}, {Rothman}, {Tan}, {Hu}, {Kassi},
  {Campargue}, \& {Medvedev}}]{li15rovibrational}
{Li}, G., {Gordon}, I.~E., {Rothman}, L.~S., {et~al.} 2015, \apjs, 216, 15,
  \dodoi{10.1088/0067-0049/216/1/15}

\bibitem[{{Lopez} \& {Fortney}(2014)}]{lopez2014}
{Lopez}, E.~D., \& {Fortney}, J.~J. 2014, \apj, 792, 1,
  \dodoi{10.1088/0004-637X/792/1/1}

\bibitem[{{Luque} \& {Pall{\'e}}(2022)}]{luque2022}
{Luque}, R., \& {Pall{\'e}}, E. 2022, Science, 377, 1211,
  \dodoi{10.1126/science.abl7164}

\bibitem[{{Luque} {et~al.}(2021){Luque}, {Serrano}, {Molaverdikhani}, {Nixon},
  {Livingston}, {Guenther}, {Pall{\'e}}, {Madhusudhan}, {Nowak}, {Korth},
  {Cochran}, {Hirano}, {Chaturvedi}, {Goffo}, {Albrecht}, {Barrag{\'a}n},
  {Brice{\~n}o}, {Cabrera}, {Charbonneau}, {Cloutier}, {Collins}, {Collins},
  {Col{\'o}n}, {Crossfield}, {Csizmadia}, {Dai}, {Deeg}, {Esposito},
  {Fridlund}, {Gandolfi}, {Georgieva}, {Glidden}, {Goeke}, {Grziwa}, {Hatzes},
  {Henze}, {Howell}, {Irwin}, {Jenkins}, {Jensen}, {K{\'a}bath}, {Kidwell},
  {Kielkopf}, {Knudstrup}, {Lam}, {Latham}, {Lissauer}, {Mann}, {Matthews},
  {Mireles}, {Narita}, {Paegert}, {Persson}, {Redfield}, {Ricker}, {Rodler},
  {Schlieder}, {Scott}, {Seager}, {{\v{S}}ubjak}, {Tan}, {Ting}, {Vanderspek},
  {Van Eylen}, {Winn}, \& {Ziegler}}]{luque2021}
{Luque}, R., {Serrano}, L.~M., {Molaverdikhani}, K., {et~al.} 2021, \aap, 645,
  A41, \dodoi{10.1051/0004-6361/202039455}

\bibitem[{{Luque} {et~al.}(2022){Luque}, {Nowak}, {Hirano}, {Kossakowski},
  {Pall{\'e}}, {Nixon}, {Morello}, {Amado}, {Albrecht}, {Caballero},
  {Cifuentes}, {Cochran}, {Deeg}, {Dreizler}, {Esparza-Borges}, {Fukui},
  {Gandolfi}, {Goffo}, {Guenther}, {Hatzes}, {Henning}, {Kabath}, {Kawauchi},
  {Korth}, {Kotani}, {Kudo}, {Kuzuhara}, {Lafarga}, {Lam}, {Livingston},
  {Morales}, {Muresan}, {Murgas}, {Narita}, {Osborne}, {Parviainen},
  {Passegger}, {Persson}, {Quirrenbach}, {Redfield}, {Reffert}, {Reiners},
  {Ribas}, {Serrano}, {Tamura}, {Van Eylen}, {Watanabe}, \& {Zapatero
  Osorio}}]{luque2022_g940b}
{Luque}, R., {Nowak}, G., {Hirano}, T., {et~al.} 2022, \aap, 666, A154,
  \dodoi{10.1051/0004-6361/202244426}

\bibitem[{{MacDonald} \& {Lewis}(2022)}]{MacDonald2022ApJ}
{MacDonald}, R.~J., \& {Lewis}, N.~K. 2022, \apj, 929, 20,
  \dodoi{10.3847/1538-4357/ac47fe}

\bibitem[{{Miller-Ricci} \& {Fortney}(2010)}]{millerricci2010}
{Miller-Ricci}, E., \& {Fortney}, J.~J. 2010, \apjl, 716, L74,
  \dodoi{10.1088/2041-8205/716/1/L74}

\bibitem[{{Miller-Ricci Kempton} {et~al.}(2012){Miller-Ricci Kempton},
  {Zahnle}, \& {Fortney}}]{millerricci2012}
{Miller-Ricci Kempton}, E., {Zahnle}, K., \& {Fortney}, J.~J. 2012, \apj, 745,
  3, \dodoi{10.1088/0004-637X/745/1/3}

\bibitem[{{Morley} {et~al.}(2013){Morley}, {Fortney}, {Kempton}, {Marley},
  {Visscher}, \& {Zahnle}}]{morley2013}
{Morley}, C.~V., {Fortney}, J.~J., {Kempton}, E. M.~R., {et~al.} 2013, \apj,
  775, 33, \dodoi{10.1088/0004-637X/775/1/33}

\bibitem[{{Morley} {et~al.}(2015){Morley}, {Fortney}, {Marley}, {Zahnle},
  {Line}, {Kempton}, {Lewis}, \& {Cahoy}}]{morley2015}
{Morley}, C.~V., {Fortney}, J.~J., {Marley}, M.~S., {et~al.} 2015, \apj, 815,
  110, \dodoi{10.1088/0004-637X/815/2/110}

\bibitem[{{Moses} {et~al.}(2013){Moses}, {Line}, {Visscher}, {Richardson},
  {Nettelmann}, {Fortney}, {Barman}, {Stevenson}, \&
  {Madhusudhan}}]{moses2013c}
{Moses}, J.~I., {Line}, M.~R., {Visscher}, C., {et~al.} 2013, \apj, 777, 34,
  \dodoi{10.1088/0004-637X/777/1/34}

\bibitem[{{Mukherjee} {et~al.}(2023){Mukherjee}, {Batalha}, {Fortney}, \&
  {Marley}}]{mukherjee2023}
{Mukherjee}, S., {Batalha}, N.~E., {Fortney}, J.~J., \& {Marley}, M.~S. 2023,
  \apj, 942, 71, \dodoi{10.3847/1538-4357/ac9f48}

\bibitem[{{Narita} {et~al.}(2013){Narita}, {Nagayama}, {Suenaga}, {Fukui},
  {Ikoma}, {Nakajima}, {Nishiyama}, \& {Tamura}}]{narita2013a}
{Narita}, N., {Nagayama}, T., {Suenaga}, T., {et~al.} 2013, \pasj, 65, 27,
  \dodoi{10.1093/pasj/65.2.27}

\bibitem[{{Nettelmann} {et~al.}(2011){Nettelmann}, {Fortney}, {Kramm}, \&
  {Redmer}}]{nettelmann2011}
{Nettelmann}, N., {Fortney}, J.~J., {Kramm}, U., \& {Redmer}, R. 2011, \apj,
  733, 2, \dodoi{10.1088/0004-637X/733/1/2}

\bibitem[{{Nixon} \& {Madhusudhan}(2021)}]{nixon2021}
{Nixon}, M.~C., \& {Madhusudhan}, N. 2021, \mnras, 505, 3414,
  \dodoi{10.1093/mnras/stab1500}

\bibitem[{Oduber {et~al.}(2021)Oduber, Calvo, Blanco-Alegre, Castro, Alves,
  Cerqueira, Lucarelli, Nava, Calzolai, Martin-Villacorta, Esteves, \&
  Fraile}]{oduber2021}
Oduber, F., Calvo, A.~I., Blanco-Alegre, C., {et~al.} 2021, Water Research,
  190, 116758, \dodoi{https://doi.org/10.1016/j.watres.2020.116758}

\bibitem[{{Ohno} \& {Okuzumi}(2018)}]{ohno2018}
{Ohno}, K., \& {Okuzumi}, S. 2018, \apj, 859, 34,
  \dodoi{10.3847/1538-4357/aabee3}

\bibitem[{{Ohno} {et~al.}(2020){Ohno}, {Okuzumi}, \& {Tazaki}}]{ohno2020agg}
{Ohno}, K., {Okuzumi}, S., \& {Tazaki}, R. 2020, \apj, 891, 131,
  \dodoi{10.3847/1538-4357/ab44bd}

\bibitem[{{Orell-Miquel} {et~al.}(2022){Orell-Miquel}, {Murgas}, {Pall{\'e}},
  {Lamp{\'o}n}, {L{\'o}pez-Puertas}, {Sanz-Forcada}, {Nagel}, {Kaminski},
  {Casasayas-Barris}, {Nortmann}, {Luque}, {Molaverdikhani}, {Sedaghati},
  {Caballero}, {Amado}, {Bergond}, {Czesla}, {Hatzes}, {Henning},
  {Khalafinejad}, {Montes}, {Morello}, {Quirrenbach}, {Reiners}, {Ribas},
  {S{\'a}nchez-L{\'o}pez}, {Schweitzer}, {Stangret}, {Yan}, \& {Zapatero
  Osorio}}]{orellmiquel2022}
{Orell-Miquel}, J., {Murgas}, F., {Pall{\'e}}, E., {et~al.} 2022, \aap, 659,
  A55, \dodoi{10.1051/0004-6361/202142455}

\bibitem[{Parmentier {et~al.}(2013)Parmentier, Showman, \&
  Lian}]{parmentier2013}
Parmentier, V., Showman, A.~P., \& Lian, Y. 2013, A\&A, 558, A91,
  \dodoi{10.1051/0004-6361/201321132}

\bibitem[{{Piette} \& {Madhusudhan}(2020)}]{piette2020}
{Piette}, A. A.~A., \& {Madhusudhan}, N. 2020, \apj, 904, 154,
  \dodoi{10.3847/1538-4357/abbfb1}

\bibitem[{{Piette} {et~al.}(2020){Piette}, {Madhusudhan}, {McKemmish},
  {Gandhi}, {Masseron}, \& {Welbanks}}]{piette2020_tio}
{Piette}, A. A.~A., {Madhusudhan}, N., {McKemmish}, L.~K., {et~al.} 2020,
  \mnras, 496, 3870, \dodoi{10.1093/mnras/staa1592}

\bibitem[{{Polyansky} {et~al.}(2018){Polyansky}, {Kyuberis}, {Zobov},
  {Tennyson}, {Yurchenko}, \& {Lodi}}]{Polyansky2018H2O}
{Polyansky}, O.~L., {Kyuberis}, A.~A., {Zobov}, N.~F., {et~al.} 2018, \mnras,
  480, 2597, \dodoi{10.1093/mnras/sty1877}

\bibitem[{{Rackham} {et~al.}(2017){Rackham}, {Espinoza}, {Apai},
  {L{\'o}pez-Morales}, {Jord{\'a}n}, {Osip}, {Lewis}, {Rodler}, {Fraine},
  {Morley}, \& {Fortney}}]{rackham2017}
{Rackham}, B., {Espinoza}, N., {Apai}, D., {et~al.} 2017, \apj, 834, 151,
  \dodoi{10.3847/1538-4357/aa4f6c}

\bibitem[{Richard {et~al.}(2012)Richard, Gordon, Rothman, Abel, Frommhold,
  Gustafsson, Hartmann, Hermans, Lafferty, Orton, Smith, \& Tran}]{richard2012}
Richard, C., Gordon, I., Rothman, L., {et~al.} 2012, Journal of Quantitative
  Spectroscopy and Radiative Transfer, 113, 1276 ,
  \dodoi{https://doi.org/10.1016/j.jqsrt.2011.11.004}

\bibitem[{{Rogers} {et~al.}(2023){Rogers}, {Schlichting}, \&
  {Owen}}]{rogers2023arXiv}
{Rogers}, J.~G., {Schlichting}, H.~E., \& {Owen}, J.~E. 2023, arXiv e-prints,
  arXiv:2301.04321, \dodoi{10.48550/arXiv.2301.04321}

\bibitem[{{Rogers} \& {Seager}(2010)}]{rogers2010gj1214b}
{Rogers}, L.~A., \& {Seager}, S. 2010, \apj, 716, 1208,
  \dodoi{10.1088/0004-637X/716/2/1208}

\bibitem[{{Rooney} {et~al.}(2022){Rooney}, {Batalha}, {Gao}, \&
  {Marley}}]{rooney2022}
{Rooney}, C.~M., {Batalha}, N.~E., {Gao}, P., \& {Marley}, M.~S. 2022, \apj,
  925, 33, \dodoi{10.3847/1538-4357/ac307a}

\bibitem[{{Rothman} {et~al.}(2010){Rothman}, {Gordon}, {Barber}, {Dothe},
  {Gamache}, {Goldman}, {Perevalov}, {Tashkun}, \& {Tennyson}}]{rothman2010}
{Rothman}, L.~S., {Gordon}, I.~E., {Barber}, R.~J., {et~al.} 2010, \jqsrt, 111,
  2139, \dodoi{10.1016/j.jqsrt.2010.05.001}

\bibitem[{Rothman {et~al.}(2010)Rothman, Gordon, Barber, Dothe, Gamache,
  Goldman, Perevalov, Tashkun, \& Tennyson}]{HITEMP2010}
Rothman, L.~S., Gordon, I.~E., Barber, R.~J., {et~al.} 2010, Journal of
  Quantitative Spectroscopy and Radiative Transfer, 111,
  \dodoi{10.1016/j.jqsrt.2010.05.001}

\bibitem[{{Rothman} {et~al.}(2013){Rothman}, {Gordon}, {Babikov}, {Barbe},
  {Chris Benner}, {Bernath}, {Birk}, {Bizzocchi}, {Boudon}, {Brown},
  {Campargue}, {Chance}, {Cohen}, {Coudert}, {Devi}, {Drouin}, {Fayt}, {Flaud},
  {Gamache}, {Harrison}, {Hartmann}, {Hill}, {Hodges}, {Jacquemart}, {Jolly},
  {Lamouroux}, {Le Roy}, {Li}, {Long}, {Lyulin}, {Mackie}, {Massie},
  {Mikhailenko}, {M{\"u}ller}, {Naumenko}, {Nikitin}, {Orphal}, {Perevalov},
  {Perrin}, {Polovtseva}, {Richard}, {Smith}, {Starikova}, {Sung}, {Tashkun},
  {Tennyson}, {Toon}, {Tyuterev}, \& {Wagner}}]{rothman2013}
{Rothman}, L.~S., {Gordon}, I.~E., {Babikov}, Y., {et~al.} 2013, \jqsrt, 130,
  4, \dodoi{10.1016/j.jqsrt.2013.07.002}

\bibitem[{{Rustamkulov} {et~al.}(2022){Rustamkulov}, {Sing}, {Liu}, \&
  {Wang}}]{rustamkulov2022}
{Rustamkulov}, Z., {Sing}, D.~K., {Liu}, R., \& {Wang}, A. 2022, \apjl, 928,
  L7, \dodoi{10.3847/2041-8213/ac5b6f}

\bibitem[{Saumon {et~al.}(2012)Saumon, Marley, Abel, Frommhold, \&
  Freedman}]{saumon2012}
Saumon, D., Marley, M.~S., Abel, M., Frommhold, L., \& Freedman, R.~S. 2012,
  \apj, 750, 74, \dodoi{10.1088/0004-637X/750/1/74}

\bibitem[{{Schlawin} {et~al.}(2023){Schlawin}, {Beatty}, {Brooks}, {Nikolov},
  {Greene}, {Espinoza}, {Glidic}, {Baka}, {Egami}, {Stansberry}, {Boyer},
  {Gennaro}, {Leisenring}, {Hilbert}, {Misselt}, {Kelly}, {Canipe}, {Beichman},
  {Correnti}, {Knight}, {Jurling}, {Perrin}, {Feinberg}, {McElwain}, {Bond},
  {Ciardi}, {Kendrew}, \& {Rieke}}]{schlawin2023}
{Schlawin}, E., {Beatty}, T., {Brooks}, B., {et~al.} 2023, \pasp, 135, 018001,
  \dodoi{10.1088/1538-3873/aca718}

\bibitem[{Southard \& Green(2018)}]{perry2018}
Southard, M., \& Green, D. 2018, Perry's Chemical Engineers' Handbook, 9th
  Edition (McGraw-Hill Education).
\newblock \url{https://books.google.com/books?id=eW5eswEACAAJ}

\bibitem[{{Spake} {et~al.}(2022){Spake}, {Oklop{\v{c}}i{\'c}}, {Hillenbrand},
  {Knutson}, {Kasper}, {Dai}, {Orell-Miquel}, {Vissapragada}, {Zhang}, \&
  {Bean}}]{spake2022}
{Spake}, J.~J., {Oklop{\v{c}}i{\'c}}, A., {Hillenbrand}, L.~A., {et~al.} 2022,
  \apjl, 939, L11, \dodoi{10.3847/2041-8213/ac88c9}

\bibitem[{{Steinrueck} {et~al.}(2021){Steinrueck}, {Showman}, {Lavvas},
  {Koskinen}, {Tan}, \& {Zhang}}]{steinrueck2021}
{Steinrueck}, M.~E., {Showman}, A.~P., {Lavvas}, P., {et~al.} 2021, \mnras,
  504, 2783, \dodoi{10.1093/mnras/stab1053}

\bibitem[{{Stevenson}(2016)}]{stevenson2016}
{Stevenson}, K.~B. 2016, \apjl, 817, L16

\bibitem[{{Stock} {et~al.}(2022){Stock}, {Kitzmann}, \& {Patzer}}]{stock2022}
{Stock}, J.~W., {Kitzmann}, D., \& {Patzer}, A. B.~C. 2022, \mnras, 517, 4070,
  \dodoi{10.1093/mnras/stac2623}

\bibitem[{{Stock} {et~al.}(2018){Stock}, {Kitzmann}, {Patzer}, \&
  {Sedlmayr}}]{stock2018}
{Stock}, J.~W., {Kitzmann}, D., {Patzer}, A. B.~C., \& {Sedlmayr}, E. 2018,
  \mnras, 479, 865, \dodoi{10.1093/mnras/sty1531}

\bibitem[{Sun \& Ariya(2006)}]{sun2006}
Sun, J., \& Ariya, P.~A. 2006, Atmospheric Environment, 40, 795,
  \dodoi{https://doi.org/10.1016/j.atmosenv.2005.05.052}

\bibitem[{{Teske} {et~al.}(2013){Teske}, {Turner}, {Mueller}, \&
  {Griffith}}]{teske2013}
{Teske}, J.~K., {Turner}, J.~D., {Mueller}, M., \& {Griffith}, C.~A. 2013,
  \mnras, 431, 1669, \dodoi{10.1093/mnras/stt286}

\bibitem[{{The JWST Transiting Exoplanet Community Early Release Science Team}
  {et~al.}(2022){The JWST Transiting Exoplanet Community Early Release Science
  Team}, {Ahrer}, {Alderson}, {Batalha}, {Batalha}, {Bean}, {Beatty}, {Bell},
  {Benneke}, {Berta-Thompson}, {Carter}, {Crossfield}, {Espinoza}, {Feinstein},
  {Fortney}, {Gibson}, {Goyal}, {Kempton}, {Kirk}, {Kreidberg},
  {L{\'o}pez-Morales}, {Line}, {Lothringer}, {Moran}, {Mukherjee}, {Ohno},
  {Parmentier}, {Piaulet}, {Rustamkulov}, {Schlawin}, {Sing}, {Stevenson},
  {Wakeford}, {Allen}, {Birkmann}, {Brande}, {Crouzet}, {Cubillos}, {Damiano},
  {D{\'e}sert}, {Gao}, {Harrington}, {Hu}, {Kendrew}, {Knutson}, {Lagage},
  {Leconte}, {Lendl}, {MacDonald}, {May}, {Miguel}, {Molaverdikhani}, {Moses},
  {Murray}, {Nehring}, {Nikolov}, {Petit dit de la Roche}, {Radica}, {Roy},
  {Stassun}, {Taylor}, {Waalkes}, {Wachiraphan}, {Welbanks}, {Wheatley},
  {Aggarwal}, {Alam}, {Banerjee}, {Barstow}, {Blecic}, {Casewell}, {Changeat},
  {Chubb}, {Col{\'o}n}, {Coulombe}, {Daylan}, {de Val-Borro}, {Decin}, {Dos
  Santos}, {Flagg}, {France}, {Fu}, {Garc{\'\i}a Mu{\~n}oz}, {Gizis},
  {Glidden}, {Grant}, {Heng}, {Henning}, {Hong}, {Inglis}, {Iro}, {Kataria},
  {Komacek}, {Krick}, {Lee}, {Lewis}, {Lillo-Box}, {Lustig-Yaeger}, {Mancini},
  {Mandell}, {Mansfield}, {Marley}, {Mikal-Evans}, {Morello}, {Nixon}, {Ortiz
  Ceballos}, {Piette}, {Powell}, {Rackham}, {Ramos-Rosado}, {Rauscher},
  {Redfield}, {Rogers}, {Roman}, {Roudier}, {Scarsdale}, {Shkolnik},
  {Southworth}, {Spake}, {E Steinrueck}, {Tan}, {Teske}, {Tremblin}, {Tsai},
  {Tucker}, {Turner}, {Valenti}, {Venot}, {Waldmann}, {Wallack}, {Zhang}, \&
  {Zieba}}]{2022arXiv220811692T}
{The JWST Transiting Exoplanet Community Early Release Science Team}, {Ahrer},
  E.-M., {Alderson}, L., {et~al.} 2022, arXiv e-prints, arXiv:2208.11692,
  \dodoi{10.48550/arXiv.2208.11692}

\bibitem[{Toon {et~al.}(1988)Toon, Turco, Westphal, Malone, \& Liu}]{toon1988}
Toon, O.~B., Turco, R.~P., Westphal, D., Malone, R., \& Liu, M.~S. 1988,
  Journal of the Atmospheric Sciences, 45, 2123

\bibitem[{{Tsai} {et~al.}(2022){Tsai}, {Lee}, {Powell}, {Gao}, {Zhang},
  {Moses}, {H{\'e}brard}, {Venot}, {Parmentier}, {Jordan}, {Hu}, {Alam},
  {Alderson}, {Batalha}, {Bean}, {Benneke}, {Bierson}, {Brady}, {Carone},
  {Carter}, {Chubb}, {Inglis}, {Leconte}, {Lopez-Morales}, {Miguel},
  {Molaverdikhani}, {Rustamkulov}, {Sing}, {Stevenson}, {Wakeford}, {Yang},
  {Aggarwal}, {Baeyens}, {Barat}, {Borro}, {Daylan}, {Fortney}, {France},
  {Goyal}, {Grant}, {Kirk}, {Kreidberg}, {Louca}, {Moran}, {Mukherjee},
  {Nasedkin}, {Ohno}, {Rackham}, {Redfield}, {Taylor}, {Tremblin}, {Visscher},
  {Wallack}, {Welbanks}, {Youngblood}, {Ahrer}, {Batalha}, {Behr},
  {Berta-Thompson}, {Blecic}, {Casewell}, {Crossfield}, {Crouzet}, {Cubillos},
  {Decin}, {D{\'e}sert}, {Feinstein}, {Gibson}, {Harrington}, {Heng},
  {Henning}, {Kempton}, {Krick}, {Lagage}, {Lendl}, {Line}, {Lothringer},
  {Mansfield}, {Mayne}, {Mikal-Evans}, {Palle}, {Schlawin}, {Shorttle},
  {Wheatley}, \& {Yurchenko}}]{tsai2022arXiv}
{Tsai}, S.-M., {Lee}, E. K.~H., {Powell}, D., {et~al.} 2022, arXiv e-prints,
  arXiv:2211.10490, \dodoi{10.48550/arXiv.2211.10490}

\bibitem[{Turco {et~al.}(1979)Turco, Hamill, Toon, Whitten, \&
  Kiang}]{turco1979}
Turco, R.~P., Hamill, P., Toon, O.~B., Whitten, R.~C., \& Kiang, C.~S. 1979,
  Journal of the Atmospheric Sciences, 36, 699

\bibitem[{{Valencia} {et~al.}(2013){Valencia}, {Guillot}, {Parmentier}, \&
  {Freedman}}]{valencia2013}
{Valencia}, D., {Guillot}, T., {Parmentier}, V., \& {Freedman}, R.~S. 2013,
  \apj, 775, 10, \dodoi{10.1088/0004-637X/775/1/10}

\bibitem[{{Van Eylen} {et~al.}(2018){Van Eylen}, {Agentoft}, {Lundkvist},
  {Kjeldsen}, {Owen}, {Fulton}, {Petigura}, \& {Snellen}}]{vaneylen2018}
{Van Eylen}, V., {Agentoft}, C., {Lundkvist}, M.~S., {et~al.} 2018, \mnras,
  479, 4786, \dodoi{10.1093/mnras/sty1783}

\bibitem[{{Wakeford} \& {Sing}(2015)}]{wakeford2015}
{Wakeford}, H.~R., \& {Sing}, D.~K. 2015, \aap, 573, A122,
  \dodoi{10.1051/0004-6361/201424207}

\bibitem[{West \& Smith(1991)}]{west1991evidence}
West, R.~A., \& Smith, P.~H. 1991, \icarus, 90, 330

\bibitem[{{Western} {et~al.}(2018){Western}, {Carter-Blatchford}, {Crozet},
  {Ross}, {Morville}, \& {Tokaryk}}]{western18}
{Western}, C.~M., {Carter-Blatchford}, L., {Crozet}, P., {et~al.} 2018, Journal
  of Quantitative Spectroscopy and Radiative Transfer, 219, 127,
  \dodoi{10.1016/j.jqsrt.2018.07.017}

\bibitem[{{Wilson} {et~al.}(2014){Wilson}, {Col{\'o}n}, {Sing}, {Ballester},
  {D{\'e}sert}, {Ehrenreich}, {Ford}, {Fortney}, {Lecavelier des Etangs},
  {L{\'o}pez-Morales}, {Morley}, {Pettitt}, {Pont}, \&
  {Vidal-Madjar}}]{wilson2014}
{Wilson}, P.~A., {Col{\'o}n}, K.~D., {Sing}, D.~K., {et~al.} 2014, \mnras, 438,
  2395, \dodoi{10.1093/mnras/stt2356}

\bibitem[{{Yu} {et~al.}(2021){Yu}, {He}, {Zhang}, {H{\"o}rst}, {Dymont},
  {McGuiggan}, {Moses}, {Lewis}, {Fortney}, {Gao}, {Kempton}, {Moran},
  {Morley}, {Powell}, {Valenti}, \& {Vuitton}}]{yu2021}
{Yu}, X., {He}, C., {Zhang}, X., {et~al.} 2021, Nature Astronomy, 5, 822,
  \dodoi{10.1038/s41550-021-01375-3}

\bibitem[{{Yurchenko} {et~al.}(2011){Yurchenko}, {Barber}, \&
  {Tennyson}}]{yurchenko2011}
{Yurchenko}, S.~N., {Barber}, R.~J., \& {Tennyson}, J. 2011, \mnras, 413, 1828,
  \dodoi{10.1111/j.1365-2966.2011.18261.x}

\bibitem[{{Yurchenko} \& {Tennyson}(2014)}]{yurchenko2014}
{Yurchenko}, S.~N., \& {Tennyson}, J. 2014, \mnras, 440, 1649,
  \dodoi{10.1093/mnras/stu326}

\bibitem[{Yurchenko \& Tennyson(2014)}]{yurchenko_2014}
Yurchenko, S.~N., \& Tennyson, J. 2014, Monthly Notices of the Royal
  Astronomical Society, 440, 1649, \dodoi{10.1093/mnras/stu326}

\bibitem[{{Yurchenko} {et~al.}(2013{\natexlab{a}}){Yurchenko}, {Tennyson},
  {Barber}, \& {Thiel}}]{yurchenko2013}
{Yurchenko}, S.~N., {Tennyson}, J., {Barber}, R.~J., \& {Thiel}, W.
  2013{\natexlab{a}}, Journal of Molecular Spectroscopy, 291, 69,
  \dodoi{10.1016/j.jms.2013.05.014}

\bibitem[{{Yurchenko} {et~al.}(2013{\natexlab{b}}){Yurchenko}, {Tennyson},
  {Barber}, \& {Thiel}}]{yurchenko13vibrational}
---. 2013{\natexlab{b}}, Journal of Molecular Spectroscopy, 291, 69,
  \dodoi{10.1016/j.jms.2013.05.014}

\bibitem[{{Zhang} \& {Showman}(2017)}]{zhang2017comp}
{Zhang}, X., \& {Showman}, A.~P. 2017, \apj, 836, 73,
  \dodoi{10.3847/1538-4357/836/1/73}

\end{thebibliography}
\bibliographystyle{aasjournal}

\end{document}